\documentclass[a4paper, 12pt]{article}
\usepackage{a4}
\usepackage{amsfonts}
\usepackage{amssymb}
\usepackage{epsfig}
\usepackage{psfig}

\author{}
\newcommand{\drawsquare}[2]{\hbox{%
\rule{#2pt}{#1pt}\hskip-#2pt
\rule{#1pt}{#2pt}\hskip-#1pt
\rule[#1pt]{#1pt}{#2pt}}\rule[#1pt]{#2pt}{#2pt}\hskip-#2pt
\rule{#2pt}{#1pt}}

\newcommand{\fund}{\raisebox{-.5pt}{\drawsquare{6.5}{0.4}}}
\newcommand{\Ysymm}{\raisebox{-.5pt}{\drawsquare{6.5}{0.4}}\hskip-0.4pt%
        \raisebox{-.5pt}{\drawsquare{6.5}{0.4}}}
\newcommand{\Yasymm}{\raisebox{-3.5pt}{\drawsquare{6.5}{0.4}}\hskip-6.9pt%
        \raisebox{3pt}{\drawsquare{6.5}{0.4}}}
\newcommand{\antifund}{\overline{\fund}}

\newcommand{\be}{\begin{equation}}
\newcommand{\ee}{\end{equation}}
\newcommand{\bea}{\begin{eqnarray}}
\newcommand{\eea}{\end{eqnarray}}
\newcommand{\mbb}{\mathbb}

\def\IR{\relax{\rm I\kern-.18em R}}

\begin{document}

\title{
\begin{flushright} \vspace{-2cm}
{\small MPP-2005-7 \\
UPR-1093-T\\
MAD-TH-05-2 \\
 \vspace{-0.35cm}
hep-th/0502005} \end{flushright} \vspace{3cm} Toward Realistic Intersecting
D-Brane Models }
\author{}
\date{}

\maketitle

\begin{center}
Ralph Blumenhagen$^1$, Mirjam Cveti\v c$^2$, Paul Langacker$^2$, and Gary Shiu$^{3,4}$\\
\vspace{0.5cm}
\emph{$^1$ Max-Planck-Institut f\"ur Physik, F\"ohringer Ring 6, \\
D-80805 M\"unchen, GERMANY}\\
\vspace{0.1cm} \emph{$^2$ Department of Physics and Astronomy, University of
Pennsylvania,
 Philadelphia, PA 19104-6396, USA }\\
\vspace{0.1cm} \emph{$^3$ Department of Physics, University of Wisconsin,
  Madison WI 53706, USA$^{\dagger}$} \\
\vspace{0.1cm} \emph{$^4$ Perimeter Institute for Theoretical Physics, \\
Waterloo, Ontario N2L 2Y5, CANADA}

\vspace{2cm}
\end{center}

\begin{abstract}
\noindent We provide a pedagogical  introduction to a
recently studied class
of phenomenologically interesting string models, 
known as Intersecting 
D-Brane Models. The gauge fields of the Standard-Model are localized  on
D-branes wrapping certain compact cycles on an underlying geometry, whose
intersections  can give rise to chiral fermions. We address the  basic issues
and also provide an overview of the recent activity in this field.
This article  is intended to serve non-experts  with
explanations of the fundamental aspects, and also to provide some
orientation for both experts and non-experts in this active field of string
phenomenology.

\end{abstract}

\vfill
\hrulefill\hspace*{4in}

{\footnotesize $^{\dagger}$Permanent address}

\thispagestyle{empty} \clearpage

\tableofcontents

\section{INTRODUCTION}
\label{sintro}

By now we have ample evidence that the Standard Model of particle physics
extended by right handed neutrinos or other mechanisms of neutrino masses
 describes nature with very high  accuracy
up to the energy range of the weak scale $E_W=10^2$ GeV. The only missing
ingredient is the Higgs particle itself, which is expected to be detected at
the Tevatron or the LHC. However, from a more formal point of view, the
Standard Model is not completely satisfactory for essentially two reasons.
First it contains 26  free parameters (not counting arbitrary electric charges)
like the masses and couplings of
fermions and bosons, which have to be measured and among  which no relation is
apparent. Second, the Standard Model is formulated as a local four-dimensional
quantum field theory and as such it does not include gravity. In fact, the
Einstein theory of general relativity cannot simply be quantized according to
the rules of local quantum field theory. Therefore, the physics we know 
of
cannot describe our universe at very high energies where quantum effects of
gravity become important.

Given these two formal shortcomings, in an ideal world we might hope that both
problems actually have the same solution. Maybe there exists a fundamental
quantum theory, which combines the Standard Model and General Relativity into
a unified framework and at the same time substantially reduces the  number of
independent parameters in the Standard Model. As this unified theory may be
geometric in nature, one might envision that some of the structure of the
Standard Model turns out to have a geometric origin as well.

We are currently in the situation that we do not know for sure what this final
theory is, but at least we have a good candidate for it, which still has to
reveal many of  its secrets. This candidate theory of quantum gravity is
called superstring theory and has been studied intensively during the last
three decades. Since superstring theory is anomaly-free only in ten space-time
dimensions, to make contact with the universe surrounding us we have to
explain what happened to the other six dimensions without contradicting
experiments. Compactifying \`a la Kaluza-Klein string theory on a compact
six-dimensional space of very tiny dimensions, our visible world would be
interpreted  as an effective four-dimensional theory, where one only keeps the
states of lowest mass. The question immediately arising is whether the formal
string equations of motion allow for six-dimensional spaces such that the
low-energy four dimensional world resembles the Standard Model of particle
physics. As a first approach, it would be too  ambitious to require that all
the couplings come out correctly. Instead, to begin with   one has to think
about stringy mechanisms for generating gauge theories with chiral matter
organized in replicated families.

The subfield of string theory concerned with such questions is called string
phenomenology and has been pursued since the mid eighties. Already at that
time is was clear that there exist two different types of ten-dimensional
superstring theories containing gauge fields on the perturbative level.
The heterotic string theories
contain only closed oriented strings and can support $SO(32)$ or $E_8\times
E_8$ gauge groups, whereas in the non-oriented Type I string theory the gauge
degrees of freedom arise from open strings, which can only support the gauge
group $SO(32)$. In Type I string theory,
the two possible orientations of the string are
identified; in other words, one is gauging the word-sheet parity
transformation.

 From the mid eighties to the mid nineties string theorists were mostly
studying $E_8\times E_8$ heterotic string compactifications, as it seemed to
be more natural to embed the Standard Model gauge group $SU(3)_C\times
SU(2)_W\times U(1)_Y$ into one of the $E_8$ factors and consider the second
$E_8$ as  a hidden gauge group which might provide the infrared physics for
supersymmetry breaking. In fact, it turned out that six-dimensional manifolds
with $SU(3)$ holonomy, so-called Calabi-Yau manifolds, also give rise to
chiral fermions, which in the most simple scenario come in identical families
where the multiplicity is given by one-half of the Euler number of the
Calabi-Yau manifold. Many examples of such Calabi-Yau manifolds were
constructed, including for instance toroidal orbifolds, hypersurfaces in
weighted projective space, or toric varieties.

In the mid nineties string theory encountered an intellectual phase transition
triggered by the realization that not only supersymmetric gauge theories but
also string theories can be related by various dualities, some of which
exchange weak and strong coupling. While before people were merely studying
perturbative aspects of string theory, now it was possible to move beyond the
perturbative framework and to catch  a glimpse of the non-perturbative physics
of string theory. The conjectured web of string dualities relied on a
speculative theory in eleven space-time dimensions, which was called M-theory.
String theorists believe that this M-theory is actually the fundamental theory,
of which the various string theories arise in certain perturbative limits.

In the process of establishing these dualities it became clear that at the
non-perturbative level string theory is not only a theory of strings but also
contains even higher dimensional objects called p-branes, which have p
space-like  and one time-like dimension. Surprisingly, the fluctuations of a
certain subset of these p-branes, so-called D-branes, are again described by a string
theory, which in this case is an open string theory with endpoints on the
brane. Since at the massless level these D-branes support gauge fields, they
are natural candidates for string phenomenology. The question is whether one
can construct consistent string compactifications with D-branes in the
background. The easiest example is the aforementioned Type I string itself,
which contains space-time filling D9-branes placed on top of the topological
defect introduced  by the gauging of the world-sheet parity. This already
indicates that for getting models with D-branes one should consider
generalizations of the Type I string. Such models, nowadays called
orientifolds, have been studied in the conformal field theory framework before
(see \cite{as02} and references therein)
and were so to say reinvented during the mid nineties from a  space-time
point of view. The aforementioned defects were called orientifold planes.

Since their  discovery many orientifold models have been constructed and there
exists an extensive literature on this subject including some review articles,
e.g.,  \cite{AD98,as02}. The present  article is not intended to be an additional
review on general orientifold models, but instead focuses on a
phenomenologically interesting class of orientifold models, which comes with
its  own intertwined  history.

The class of models covered here has its origin in the observation that two
generically intersecting D-branes can support chiral fermions on the
intersection locus  \cite{bdl96}. Therefore, one is led to models which not
only contain D-branes on top of or parallel to orientifold planes, but one
should also allow these D-branes to be placed such that there exist chiral
intersections  as long as they do not violate the stringy consistency
conditions. Historically, the first models of this kind were discussed in a
T-dual formulation with magnetic fluxes in Type I string theory by C. Bachas
\cite{CB95}. Providing  the complete stringy picture of this early idea  and
showing its dual formulation in terms of intersecting D-branes, the first
really intersecting D-brane  models were constructed  in
\cite{bgkl00a,bgkl00b}. Independently, supersymmetric compactifications of the
Type I string to six-dimensions with magnetic fluxes were discussed in
\cite{aads00,as00a}. Intersecting branes and magnetized branes are equivalent
descriptions \cite{bdl96}. Therefore  without loss of generality, in this
review article we stick to the more intuitive picture of geometrically
intersecting D-branes. Non-chiral orientifold models with  D-branes
intersecting at angles had been considered even before
\cite{bgk99,GP99,ab99,bgk99a,fhs00,bcs04}.

To emphasize it again,
intersecting branes provide a stringy mechanism for
generating not only gauge symmetries  but also chiral fermions, where family replication is
achieved by multiple topological intersection numbers of various D-branes.
Therefore, these models
provide a beautiful geometric picture of some of the fundamental   ingredients
of the Standard Model.

After the introduction of these kinds of models, some generalizations and
additional profound issues were discussed in \cite{afiru00,afiru00a,bkl00}. In
the original models of \cite{bgkl00a} the closed string background was simply
a flat torus, for which it could be shown that flat non-trivially intersecting
D-branes always break supersymmetry explicitly at the string scale. Therefore,
chiral models were necessarily non-supersymmetric. For a field theorist this
is not a problem, as the Standard Model as we know  is non-supersymmetric
anyway. However, from the stringy point of view, supersymmetry is generally
the mechanism which guarantees that string compactifications are stable. 
In order for a string vacuum to have a life-time longer than the Planck (or
better string) time, it seems desirable to start with a supersymmetric vacuum and
then break supersymmetry softly in a controlled way. Though many papers in the literature 
deal with non-supersymmetric intersecting D-brane  models, the reader should keep in 
mind that for these models, even though the open string sectors look amazingly similar to the
Standard Model \cite{imr01}, one generically encounters stability problems in the closed
string sector \cite{bklo01}. Due to their popularity and some issues which carry over to the
supersymmetric models, we will also cover the non-supersymmetric models in
this review, but our main focus will be on chiral supersymmetric models, first constructed
in \cite{csu01,csu01a}. For them,  Standard-like Models are much harder to construct 
and one has to consider more general than purely toroidal backgrounds, e.g., orbifolds.

In the original setting, one considered orientifolds of Type IIA string
theory, which contain only orientifold six-planes, whose charges are canceled
by introducing intersecting D6-branes. Such models can be defined on general
six-dimensional manifolds,  where the requirement of supersymmetry however
implies this to be a Calabi-Yau  manifold. Various generalizations with
D-branes of other dimensions have been contemplated, but we think that the
original models are the most natural class of intersecting D-brane models (as
for instance they are related to M-theory compactifications on $G_2$
manifolds). Therefore, throughout this article we will mainly work in this
framework and only mention the possible generalizations.

Different aspects of these intersecting D-brane models have been discussed
during the last four years  in a large number of papers,  which can be mainly
categorized into three classes (we will provide the references in the
appropriate sections of the main text).  First, there are the stringy model
building aspects, which in particular include the derivation of the stringy
consistency conditions (R-R tadpole cancellation conditions) and the
computation of the massless spectrum. Second, tools have been developed to
compute for a given string model the four-dimensional low-energy effective
action, which includes tree level expressions for Yukawa couplings, higher
point correlation functions,  gauge couplings, Fayet-Iliopoulos terms and
K\"ahler potentials. Moreover, for the gauge couplings also one-loop
corrections have been computed. This program of determining the low-energy
effective actionon-chiral states  can typically obtain  a string scale mass
after deformations of the brane configuration. It is not complete yet and 
has
mainly been applied to purely toroidal (orbifold) string backgrounds. Finally,
using the results about  the effective action, people  discussed the
phenomenological low energy implications of intersecting D-brane models; some
of them turn out to be rather model independent whereas others are not and
might be used to discard certain models for phenomenological reasons. These
are the three main aspects, but of course there exist relations of
intersecting D-brane models to other branches of recent research such as
M-theory compactification on $G_2$ manifolds or compactifications with
non-trivial background fluxes. These latter developments will also be covered
in this article, where, however, we do not provide a general introduction to
$G_2$ manifolds or flux compactifications, as this would fill another review
article. Another possible connection is to the phenomenological brane world
ideas associated with possible large extra dimensions \cite{IA90,add98,aadd98} that have
been popular in recent years. While most intersecting D-brane constructions
involve only small extra dimensions (within a few orders of magnitude of  the
inverse Planck scale), it is possible (and probably necessary for
non-supersymmetric constructions) to consider internal spaces with large
dimensions, providing a stringy realization of those ideas.

The aim of this article is twofold. First it is intended to give a pedagogical
introduction to the subject and to provide the main technical tools for the
construction of intersecting D-brane models. It should allow
non-experts to understand the main aspects of the subject and enable students
to get started in this field. Second, we attempt to give as broad an overview
 as possible of developments in the field and to point out open
questions. Of course, to be as complete as possible we had to neglect many
details, and we are aware that the topics we put special emphasis on reflect
in some way our own preferences. We apologize to all those authors who feel
that their work has not been covered to a degree they believe  it deserves.
Several  articles of review type with slightly different
emphasis have appeared during the last years \cite{AU03,FM03,OTT03,
EK03, LG04, DL04,RB04}.

\section{ORIENTIFOLDS WITH INTERSECTING D-BRANES}
\label{soib} Throughout this technical introduction into intersecting D-brane
models we assume that the reader is familiar at least at a textbook level
\cite{gsw87,gsw87a,JP98, JP98a,CJ03,BZ04} with the basic notions of string
theory including the concept of D-branes.

String compactifications from ten to four space-time dimensions have been
studied throughout the history of string theory, but in the mid nineties the
second string theory revolution provided new insights into the constructions of
four-dimensional
vacua from M-theory.
As with all the progress made during this exciting
epoch, this had to do with the realization that string theory is not only a
theory of either closed or open strings but also contains in its
non-perturbative sector extended objects of  higher dimensions, so called
D-branes (see \cite{pcj96,JP96, CJ03} for reviews on D-branes). These 
D-branes are
charged under some of the massless fields appearing in the Ramond-Ramond (R-R)
sector of the ten-dimensional Type IIA/B string theories. More concretely, a
p-brane is an extended object with p space-like directions and one time-like
direction and it couples to a $(p+1)$ form potential $A_{p+1}$  as follows:
\begin{equation}
            {\cal S}_{p} =    Q_p   \int_{D_p} A_{p+1},
\end{equation}
where the integral is over the $(p+1)$ dimensional world-volume of the D-brane
and $Q_p$ denotes its  R-R charge. 
For BPS D-branes in Type IIA string theory $p$ is an even
number and in Type IIB an odd one. Polchinski was the first to realize that
the fluctuations of such D-branes can by themselves be described by a string
theory \cite{JP95}, which in this case are open strings attached to the
D-brane, i.e., with Dirichlet boundary conditions transversal to the D-brane
and Neumann boundary conditions along the D-brane \bea
         \mu=0, \ldots , p &&\quad \quad \partial_{\sigma} X^{\mu}\vert_{\sigma=0,\pi}=0, \nonumber \\
        \mu=p+1, \ldots , 9 && \quad \quad \partial_{\tau} X^{\mu}\vert_{\sigma=0,\pi}=0.
\eea
where $(\sigma,\tau)$ denote the world-sheet space and time coordinates and
$X^\mu$ the space-time coordinates. Their  world-sheet superpartners
are denoted as $\psi^\mu$ in the following.
Upon quantization of an open string, the  massless excitations
$\psi^\mu_{-{1\over 2}}|0\rangle$  give rise to  a $U(1)$ gauge field, which
can only have momentum along the D-brane and is therefore confined to it. It
is precisely the occurrence of these gauge fields which makes
D-branes interesting objects for string model building. If one can construct
string models with D-branes in the background, then one has a natural source
of gauge fields, which are of fundamental importance   in the Standard
Model of particle physics. Placing $N$ D-branes on top of each
other the gauge fields on the branes transform in the adjoint representation of
the gauge group $U(N)$.

\medskip

In this section, we will discuss the general rules for constructing
intersecting D-brane models. In subsection \ref{ssss}, employing the effective
gauge and gravitational couplings, we discuss how the
string scale of the intersecting D-brane models depends on the closed string
moduli. Then in subsection \ref{ssc}, we discuss how chiral
fermions arise at the intersection of D-branes. The fact that D-branes can
intersect more than once in a compact space gives rise to the  interesting
feature of family replication, which will be discussed in subsection
\ref{ssfr}. In addition to R-R charges, D-branes also couple
gravitationally which means that they have tension.
To cancel the positive contribution to the vacuum energy from the tension of
D-branes, we need to introduce negative tension objects known as orientifold
planes. The notion of orientifolds will be discussed in subsection \ref{sso}.
The total R-R charge carried by the D-branes and orientifold planes has to
vanish for consistency. Such tadpole cancellation conditions are derived in
subsection \ref{ssrrtc}. With the configuration of D-branes and orientifold
planes that satisfy the tadpole conditions, one can derive the spectrum of
massless open strings ending on the D-branes. The chiral part of the spectrum
is summarized in subsection \ref{ssms}. In general, there are anomalous
$U(1)$'s in intersecting D-brane models whose anomalies are canceled by the
generalized Green-Schwarz mechanism as explained in subsection \ref{ssggsm}.
In subsection \ref{sssb}, we discuss the conditions for the configuration of
D6-branes to be supersymmetric. It turns out that they have to wrap around
three-cycles known as  special Lagrangian cycles. Interestingly, the
intersecting D6-brane models which preserve ${\cal N}=1$ supersymmetry in four
dimensions can be lifted to eleven-dimensional M theory as compactifications
on singular $G_2$ manifolds. The lift and the connection to how chiral
fermions arise in the $G_2$ context are discussed in \ref{ssg2}. As two warmup
examples for later use,  intersecting D-branes on $T^6$ and the $T^6/\mbb{Z}_2\times \mbb{Z}_2$
orbifold    are presented in subsection \ref{sse}.

\subsection{The String Scale}
\label{ssss} The localization of gauge fields on D-branes provides a concrete
stringy realization of the brane world scenario in which the Standard Model
fields are confined on the branes whereas gravity propagates in the bulk. As a
result, the four-dimensional gauge couplings are determined by the volume of
the cycles that the D-branes wrap around, while the gravitational coupling
depends on the total internal volume. This opens up the possibility of
lowering the string scale.
 More specifically, by dimensional reduction to four dimensions\footnote{The factors of $2 \pi$ were carefully worked out in \cite{st98}.}:
\begin{eqnarray}
\label{gauge-gravitational-couplings}
\frac{1}{g_{YM}^2} &=& \frac{M_s^{p-3} V_{p-3}}{(2 \pi)^{p-2} g_s} \nonumber \\
M_P^2 &=& \frac{M_s^8 V_6}{(2 \pi)^7 g_s^2},
\end{eqnarray}
where $V_{p-3}$ is the volume of the $p-3$ cycle wrapped by a  Dp-brane
(which is in general different for different branes) and $V_6$ is the
total internal volume. In this article, we will focus on models with
intersecting D6-branes so that
\begin{equation}
g_{YM}^2 M_P
= \sqrt{2 \pi} M_s \frac{\sqrt{V_6}}{V_3}.
\end{equation}
The experimental bounds on the masses of Kaluza-Klein replicas of the Standard
Model gauge bosons imply that the volume of three-cycles cannot be larger than
the inverse TeV scale generically. For a general internal space (such as a
Calabi-Yau manifold), the volumes of the three-cycles are not directly
constrained by the scale of the total internal volume, and can be much smaller
than $\sqrt{V_6}$. In this case, a large Planck mass can be generated from a
large total internal volume.
This is precisely the idea of the large extra dimension
scenario.

However, for intersecting D6-brane models in toroidal
backgrounds, $V_3$ is of the same order as $\sqrt{V_6}$ (since for chiral
models, there is no dimension transverse to all the branes) so the string
scale is of the order of the Planck scale $M_P$. There is, however, more
freedom than in theories with only closed strings (e.g., the heterotic
string), and this could be used to lower the string scale to, e.g., $10^{16}$
GeV, a certainly desirable choice for Grand Unified  Models.

\subsection{Chirality}
\label{ssc} One of the main features of the Standard Model is that the light
fermionic matter fields appear in chiral representations of the $SU(3)_C\times
SU(2)_W \times U(1)_Y$ gauge symmetry such that all gauge anomalies are
canceled. Considering just parallel D-branes in flat space one does not get
chiral matter on the branes, so that one has to invoke an additional
mechanism to realize  this phenomenologically very important feature.
Essentially, so far two ways have been proposed to realize  chirality for the
D-brane matter spectrum.  The first one is to place the D-brane not in flat space, but on so-called orbifold 
(or conifold) singularities and the second is to let the D-branes intersect at
non-trivial angles \cite{bdl96}. We will discuss here only the second 
mechanism 
in more detail and refer the interested reader for the first mechanism 
to the existing 
literature (see for instance \cite{as02,AU00} and references therein).

\begin{figure}
\epsfbox{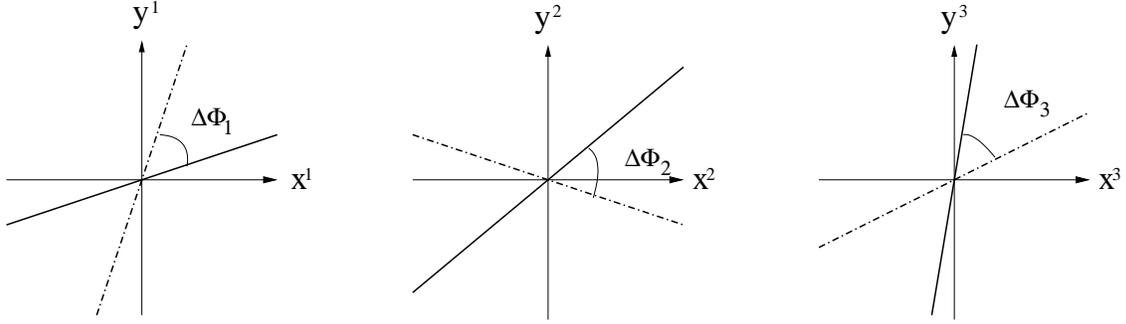} \caption{Intersecting D6-branes} \label{fib}
\end{figure}

To be more precise consider two D6-branes sharing  the  four dimensional
Minkowskian space-time. This means that in the six dimensional transversal space
the branes are three-dimensional and  wrap a three dimensional cycle.
In general position two such branes do intersect in a point in the internal
space.
Consider the simple case of a flat six-dimensional internal space.
Choosing light cone gauge, let us introduce
complex coordinates $z^i=x^{i}+iy^{i}$ with $i=0,\ldots,3$.
Then two D6-branes cover the $z^0$ plane and intersect in the other
directions as
shown in Figure \ref{fib}.

\begin{figure}
\begin{center}
\epsfbox{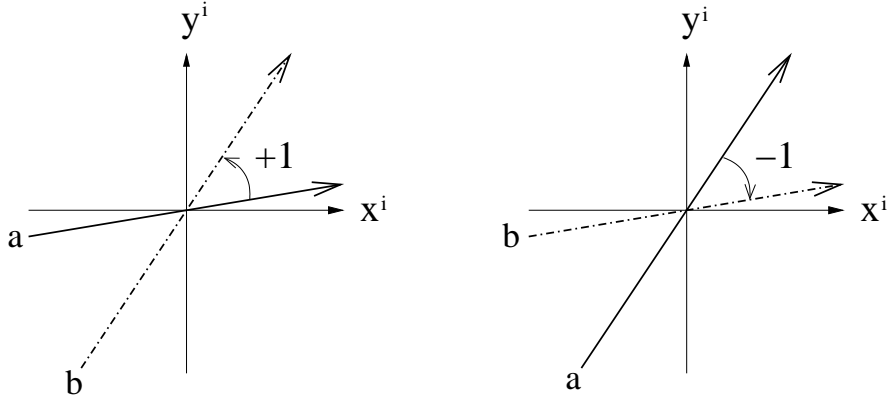} \caption{Oriented intersection} \label{foi}
\end{center}
\end{figure}

Placing for convenience one D-brane along the $x^i$ axes, an open string
stretched between two intersecting D-branes has the following boundary
conditions \bea
      \sigma=0:&&\quad\quad \partial_{\sigma} X^i = \partial_{\tau} Y^i=0 \nonumber \\
      \sigma=\pi:&&\quad\quad\phantom{-}\,  \partial_{\sigma} X^i + \tan (\Delta\Phi_i)\,  \partial_{\sigma} Y^i= 0 \\
             &&\quad\quad - \tan (\Delta\Phi_i)\, \partial_{\tau} X^i + \partial_{\tau} Y^i= 0  \nonumber
\eea
which in complex coordinates read
\bea
      \sigma=0:&&\quad\quad \partial_{\sigma} (Z^i+\overline{Z}^i)  =
                     \partial_{\tau} (Z^i-\overline{Z}^i) =0 \nonumber \\
      \sigma=\pi:&&\quad\quad \partial_{\sigma} Z^i + e^{2i \Delta\Phi_i}
                       \partial_{\sigma} \overline{Z}^i= 0 \\
                 &&\quad\quad  \partial_{\tau} Z^i  -
             e^{2i \Delta\Phi_i} \partial_{\tau} \overline{Z}^i= 0. \nonumber
\eea
Now, implementing these boundary conditions in  the mode expansion
of the fields $Z^i$ and $\overline{Z}^i$,
one finds \cite{bdl96}
\be
   Z^i(\sigma,\tau) = \sum_{n\in \mbb{Z} } {1\over (n+\epsilon_i)}\, \alpha^i_{n+\epsilon_i}\,
               e^{-i(n+\epsilon_i)(\tau+\sigma)} +
            \sum_{n\in \mbb{Z} } {1\over (n-\epsilon_i)}\, \tilde\alpha^i_{n-\epsilon_i}\,
               e^{-i(n-\epsilon_i)(\tau-\sigma)}
\ee
with $\epsilon_i=\Delta\Phi_i/\pi$
for $i\in\{1,2,3\}$.
Therefore the bosonic oscillator modes of the fields $Z^1, \ldots, Z^3$ are given by
\be \alpha^i_{n+\epsilon_i},\quad \tilde\alpha^i_{n-\epsilon_i}
\label{eom}
\ee
Similarly, for the
world-sheet fermions the modes are $\psi^i_{n+\epsilon_i}$ and
$\tilde\psi^i_{n-\epsilon_i}$ in the R-R  sector and with
an additional $1/2$-shift in the Neveu-Schwarz Neveu-Schwarz (NS-NS) sector.
Therefore, in analogy to the closed string sector, an open string between two
intersecting D-branes can be considered as a twisted open string. As a
consequence for all $\epsilon_i$ non-vanishing there are only two zero
modes in the R-R sector, $\psi^1_0, {\tilde\psi_0}^1$, which give
rise to a twofold degenerate R-R ground state. The GSO projection
eliminates one half of these states, so that one is left with only one
fermionic degree of freedom. Taking into account also the open string  with
the opposite orientation between the two D6-branes, one finally gets two
fermionic degrees of freedom corresponding to one chiral Weyl-fermion from the
four-dimensional space-time point of view. To summarize, we have found that
two generically intersecting D6-branes give rise to one chiral fermion at the
intersection point. If we now consider the intersection between a stack of $M$
D6-branes with another stack of $N$ D6-branes it is clear that the for
instance left-handed chiral
fermion transforms in the bi-fundamental representation of the $U(M)\times U(N)$
gauge symmetry.
We choose the convention that this is the $(\overline M, N)$ representation
of the gauge group. As such this result is not invariant under the
exchange of the role of
$M$ and $N$. This can be remedied by giving  an orientation to the branes and
by assigning a sign to the intersection on each plane as shown in Figure
\ref{foi}. A negative intersection simply means that one gets a left-handed
chiral fermion transforming in the conjugate representation of the gauge group.
The intersection  defined this way is anti-symmetric under exchange of
the two branes.

\subsection{Family replication}
\label{ssfr} In the last section we have seen that intersecting D-branes can
be a source for chiral fermions, which makes them very interesting candidates
for model building. However, chiral fermions in the Standard Model come in
three families  differing only by their mass scale. Therefore, it is important
to search for  a mechanism for family replication. As we will see by
considering intersecting branes on compact backgrounds such a mechanism
automatically arises.

\begin{figure}
\epsfbox{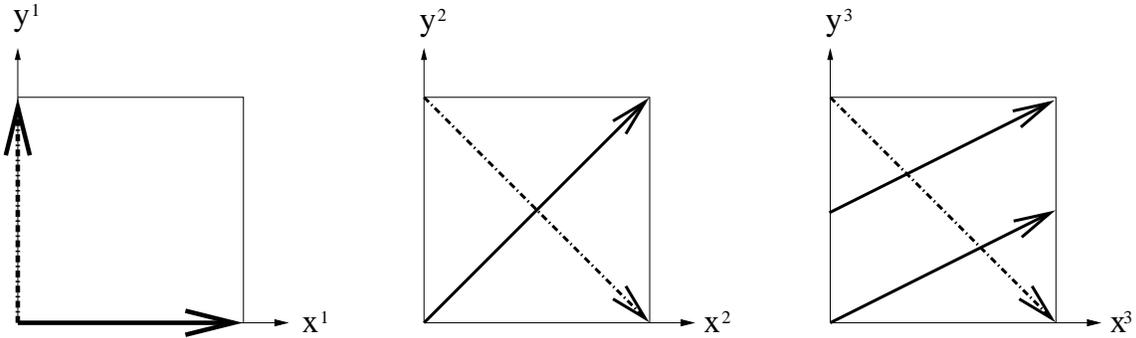} \caption{Intersecting D6-branes on a torus} \label{fibt}
\end{figure}

In the non-compact flat background depicted in Figure \ref{fib} it is clear
that the intersection number can only be $\pm 1$. However, in the compact case
like for instance a torus, it can be easily seen that the intersection number
can be larger than one. Assuming for simplicity that the background is a
six-dimensional torus with complex structure chosen such that it can be
written as $T^6=T^2\times T^2\times T^2$, a large class of D6-branes cover
only a one-dimensional cycle on each factor $T^2$. Such D6-branes have been
called factorizable in the literature and are described by three pairs of
wrapping  numbers $(n^i,m^i)$  along the fundamental 1-cycles of three $T^2$s.
In Figure \ref{fibt} we have shown two such wrapped D6-branes with wrapping
numbers $(1,0)(1,1)(2,1)$ for the first D-brane and $(0,1)(1,-1)(1,-1)$ for
the second one.

 From the picture one reads off that the intersection number  between the two
D6-branes is $I_{ab}=6$ which is just one simple example of the general
expression for the intersection number \be
     I_{ab}=\prod_{i=1}^3 (n^i_a\, m^i_b- m^i_a\, n^i_b).
\ee By deforming the D6-branes it is clear that one can easily generate
additional intersections, but they always come in pairs with positive and
negative sign, so that the net number of chiral fermions remains constant.
Therefore, what really counts the net number of chiral fermions is the
topological intersection number, which only depends on the homology classes of
the two branes.  In our case, the homological three-cycles are simply products
of three one-cycles, where the homological one-cycles on each $T^2$
are characterized just by the wrapping numbers $(n_a,m_a)$.

Generalizing the set-up we have introduced so far,
we consider compactifications of Type IIA string theory on a six-dimensional
manifold ${\cal M}$. To preserve ${\cal N}=1$ supersymmetry, the
manifold ${\cal M}$ is a Calabi-Yau manifold.
As a topological space, ${\cal M}$ has homological three-cycles $\pi_a$, $a\in\{1,\ldots,K\}$,  on
which we can wrap $N_a$ D6-branes. From the effective four-dimensional point
of view, we obtain gauge fields of $\prod_{a=1}^K U(N_a)$ localized on the seven-dimensional
world-volume of the D6-branes. Additionally, one gets chiral fermions
localized on the four-dimensional intersection locus of two branes which come
with multiplicity given by the topological intersection number $\pi_a\circ
\pi_b$ and transform in the  $(\overline{N}_a, N_b)$ representation of the
gauge group.

As we will discuss in the following sections, tadpole cancellation and
supersymmetry impose certain constraints on the three-cycles the D6-branes are
wrapped around.

\subsection{Orientifolds}
\label{sso}
As has been pointed out in  \cite{bklo01},
 non-supersymmetric models, though easy to handle, 
are unstable in the sense that their perturbative scalar
potential, which is due to the so-called NS-NS tadpoles,
gives rise to runaway behavior for many of the closed string moduli fields including
the dilaton. In
addition, for constructions based on toroidal-type  compactifications the
volume of three cycles is of the same order of magnitude as the square root of
the volume of the internal space. Thus, for the case of  intersecting
D6-branes there is no direction in the internal space that can be taken large
compared to the Planck radius while  keeping the  correct values of  gauge
couplings and the Planck scale in four-dimensions (see
eqs.(\ref{gauge-gravitational-couplings}) in section \ref{ssss}.). Therefore,
in this case  the string scale cannot be much below  the Planck scale and
thus both the  NS-NS tadpole contributions to the
potential as well as radiative corrections in  the effective theory are large,
i.e., of the order of the Planck scale.  There is a chance that
NS-NS tadpoles might be stabilized by non-perturbative
effects or by turning on fluxes, but nevertheless, in order to stay on firm
ground from the string theory perspective, we prefer to mainly consider
supersymmetric models.

The set-up introduced so far with intersecting D6-branes in Type IIA
compactifications always breaks supersymmetry. This can be seen as follows.
For a globally supersymmetric background the vacuum energy has to vanish.
However, all D6-branes have a positive contribution to the vacuum energy, as
their tension is always positive and therefore they break supersymmetry. The
only way to finally find non-trivial supersymmetric models is by introducing
objects of negative tension into the theory. It is well known that such
objects exist in string theory and that they naturally occur in so-called
orientifold models.

An orientifold is the quotient of Type II string theory  by a discrete
symmetry group $G$ including  the world-sheet parity transformation
$\Omega:(\sigma,\tau)\to (-\sigma,\tau)$. As a consequence the resulting
string models  contain non-oriented strings and their perturbative expansion
also involves non-oriented surfaces like the Klein-bottle.
Dividing out by
such a symmetry, new objects called orientifold planes arise, whose
presence can be detected for instance by computing the Klein-bottle amplitude
\be
      K=\int_0^\infty {dt\over t} {\rm Tr}\left( {\Omega\over 2} \,
                 e^{-2\pi t (L_0 +\overline L_0)}\right).
\ee
These objects, though non-dynamical, do
couple to the closed string modes and in particular
they carry tension and charge under some of the R-R fields. In other words,
there exist non-vanishing tadpoles of the closed modes on the orientifold
planes  which, as it turns out, can  have opposite sign than the corresponding terms
for D-branes.  Since the overall charge one puts on a compact space has to vanish by
Gauss' law, the contribution from the orientifold planes and the D-branes have to cancel.
We would like to emphasize that for orientifolds the
presence  of D-branes in the background is in most cases not an option
but a necessity.

\begin{figure}
\begin{center}
\epsfbox{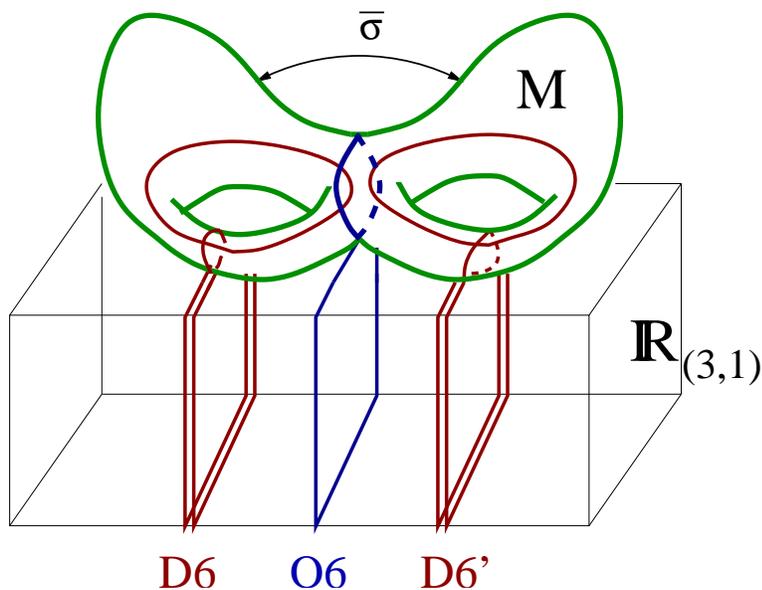} \caption{Schematic image of an $\Omega\overline\sigma (-1)^{F_L}$
orientifold with O6-planes and intersecting D6-branes. In reality
the O6-plane and the $D6$-branes would cover the entire 
flat Minkowski space.}
\label{foib}
\end{center}
\end{figure}

Which are the appropriate  orientifolds to consider so that intersecting
D6-branes might cancel the tadpoles? Clearly  we need O6-planes, meaning
that  the world-sheet parity has to be dressed with an involution, 
locally reflecting three out
of the six internal coordinates, and being a symmetry of the internal space. 
Let us assume that ${\cal M}$ admits  a complex
structure so that we locally can  introduce complex coordinates $z^i$. Now, we consider
Type IIA string theory divided out by $\Omega\overline\sigma (-1)^{F_L}$,
where $F_L$ denotes the left-moving space-time fermion number
\footnote{Note that the $(-1)^{F_L}$ factor was not explicitly written
down in many of the papers on intersecting D-brane models.}
and $\overline\sigma$  an isometric anti-holomorphic involution of ${\cal M}$.
This acts on the K\"ahler class $J$ and the holomorphic covariantly constant
three-form $\Omega_3$ as
\be
   \overline\sigma\, J = -J, \quad\quad
    \overline\sigma\,  \Omega_3 = e^{2i\varphi} \overline{
     \Omega}_3
\ee
with $\varphi\in \IR $.
For $\varphi=0$ in local coordinates this can be thought of as complex conjugation. As a
result we get an orientifold O6-plane localized at the fixed point locus of
$\overline\sigma$, which topologically is a three-cycle $\pi_{O6}$ in
$H_3({\cal M},\mbb{Z})$.  To cancel the resulting massless tadpoles we
introduce appropriate  configurations of intersecting D6-branes wrapping
homological three-cycles $\pi_a$. For  $\overline\sigma$ to be a symmetry of
the brane configuration, one also needs to wrap D6-branes on  the
$\overline\sigma$ image three-cycles $\pi'_a$. As a new feature, in
orientifold models it is also possible to get orthogonal and symplectic gauge
symmetries. The rule is very simple. If a three-cycle is invariant under the
anti-holomorphic involution one gets either $SO(2N_a)$ or $SP(2N_a)$ gauge
symmetry; if the cycle  is not-invariant one gets $U(N_a)$. Figure \ref{foib}
depicts in a simplified way the set-up discussed in this section.

\subsection{R-R Tadpole cancellation}
\label{ssrrtc} As we have mentioned already tadpole cancellation provides some
constraints on  the positions of the O6-planes  and D6-branes, which we
now  summarize. Historically, for deriving the tadpole cancellation conditions
one used an indirect method by first computing, using conformal field theory
techniques, the one-loop Klein-bottle, annulus and M\"obius strip diagrams,
and extracting from the corresponding tree-channel amplitudes
the infrared divergences due to massless tadpoles. Employing a
direct method using the Dirac Born Infeld action, here we essentially follow
\cite{bbkl02,bbkl02a}, where also more details of the derivation can be found.

Consider the  part of the supergravity Lagrangian where the R-R field $C_{7}$
appears \bea {\cal S}&=& -{1\over 4\kappa^2} \int_{\IR^{3,1}\times {\cal M}}
{ dC_{7} \wedge \star dC_{7} } +\mu_6
\sum_a{ N_a \int_{\IR^{3,1}\times\pi_a} C_{7} } \\
&+&\mu_6 \sum_a{ N_a \int_{\IR^{3,1}\times\pi'_a} C_{7} } -4 \mu_6  \int_{
\IR^{3,1}\times\pi_{{\rm O}6}  } C_{7},
\eea
where the ten-dimensional
gravitational coupling is $\kappa^2={1\over 2} (2\pi)^7(\alpha')^4$ and the
R-R charge of a D6-brane reads $\mu_6=(\alpha')^{-{7\over 2}}/(2\pi )^6$.
Note that here we have assumed that the orientifold planes are
of type $O^{(-,-)}$, i.e., they carry negative tension and R-R charge.
Recall that D-branes in this convention carry positive tension and R-R charge.
Such models have also been called orientifolds without vector
structure.
The
resulting equation of motion for the R-R field strength $G_{8}=dC_{7}$ is \be
\label{tadd} {1\over \kappa^2}\,
     d\star G_{8}=\mu_6\sum_a N_a\, \delta(\pi_a)+
                    \mu_6\sum_a N_a\, \delta(\pi'_a)
        -4 \mu_6\,  \delta(\pi_{{\rm O}6}),
\ee where $\delta(\pi_a)$ denotes the Poincar\'e dual three-form of $\pi_a$.
Since the left hand side in eq. (\ref{tadd})  is exact, the R-R tadpole
cancellation condition boils down to just a simple condition on the homology
classes \be \label{tadpole} \sum_a  N_a\, (\pi_a + \pi'_a) -4 \pi_{O6}=0. \ee
The above condition implies that the overall three-cycle all the D-branes and
orientifold planes wrap is trivial in homology. This is a restrictive
condition but it is moderate enough to admit non-trivial solutions with
branes are not simply placed right on top of the orientifold plane. 
Note that
so far we have not assumed supersymmetry and that therefore eq.
(\ref{tadpole}) does not automatically guarantee the NS-NS tadpoles to be
canceled as well.

However, it is important to note that 
the above method using the
Dirac-Born-Infeld action together with supergravity 
does not take into account {\it all} the
R-R charges carried by the D-branes.
The reason is that
D-brane charges are classifed by K-theory groups
rather than homology groups, and the R-R 
fields in general 
are not simply $p$-forms like above (see, e.g., \cite{Witten98} for a 
more detailed discussion). 
Indeed, as pointed out in \cite{Uranga00}, the cancellation of homological
R-R charges (i.e., the conditions 
(\ref{tadpole}) above) 
are not sufficient to ensure that all the R-R tadpoles vanish
and hence the consistency of the models.
The inconsistencies due to uncanceled K-theory charges
would show up as discrete global anomalies \cite{EW82}
either in the low energy spectrum or on the world-volume of a probe
D-brane \cite{Uranga00}.
One way to heuristically derive these constraints is to introduce probe 
D-branes with an 
$Sp(2n)$ gauge group, and require that the total number of  fundamental 
representations in their world-volume theory to be even.
These K-theory constraints are widely unnoticed in the model building 
literature because for simple models, they are automatically satisfied.
However, these consistency constraints 
are far from trivial. For example, such K-theory constraints for
 the $\mbb{Z}_2 \times \mbb{Z}_2$
orientifold were derived in \cite{ms04,ms04a} and have shown to 
play an important role in the construction of more realistic models.
We will discuss such constraints in more detail in subsection \ref{sse}.

\subsection{The massless spectrum}
\label{ssms}

For model building purposes it is very important to have control over the
massless spectrum arising from any kind of string compactification. For the
orientifold models with intersecting D6-branes the chiral spectrum arising
from the various open string sectors can be determined just from the
intersection numbers of the three-cycles the D6-branes are wrapped around. For
simplicity let us assume that all D6-branes wrap three-cycles not invariant
under the anti-holomorphic involution, so that the gauge symmetry is $\prod_a
U(N_a)$.  For this case the general rule for determining the massless
left-handed chiral spectrum is presented in Table \ref{tcs}.
\begin{table}
\caption{Chiral spectrum for intersecting D6-branes} \centering \vspace{3mm}
\label{tcs}
\begin{tabular}{|c|c|}
\hline
Representation  & Multiplicity \\
\hline $\Yasymm_a$
 & ${1\over 2}\left(\pi'_a\circ \pi_a+\pi_{{\rm O}6}
\circ \pi_a\right)$  \\
$\Ysymm_a$
     & ${1\over 2}\left(\pi'_a\circ \pi_a-\pi_{{\rm O}6} \circ \pi_a\right)$   \\
$(\antifund_a,\fund_b)$
 & $\pi_a\circ \pi_{b}$   \\
 $(\fund_a,\fund_b)$
 & $\pi'_a\circ \pi_{b}$
\\
\hline
\end{tabular}
\end{table}
Open strings stretched between a D-brane and  its $\o\sigma$ image 
are the only ones left invariant under the combined
operation $\Omega\overline\sigma (-1)^{F_L}$. Therefore, they  transform
in the antisymmetric or symmetric representation of the gauge group,
indicating that  the price we have to pay by
considering intersecting D-branes in an orientifold background is that more
general representations are possible for the chiral fermions. Sometimes this
is an advantage, like for constructing $SU(5)$ Grand Unified Models, but 
sometimes the
absence of such fermions imposes new conditions on the possible D-brane
set-ups.

The rule for the chiral spectrum in Table \ref{tcs} is completely general and,
as was demonstrated in \cite{bbkl02}, the chiral massless spectra from many
orientifold models discussed using conformal field theory methods in  the
existing literature can be understood  in this framework.

Moreover, one can easily check that the R-R tadpole cancellation condition
(\ref{tadpole}) together with Table \ref{tcs}  guarantees the absence of non-Abelian gauge
anomalies. Naively, there exist  Abelian and  mixed Abelian, non-Abelian
anomalies, as well as gravitational anomalies. However,  we shall see in the
subsequent section that all of these are canceled by a generalized
Green-Schwarz mechanism.

To apply Table \ref{tcs}  to concrete models, one has to compute the
intersection numbers of three-cycles, which by itself is in general not an
easy task. However, there exist backgrounds for which  generic rules can be
presented. Besides the simplest case of just a torus $T^6$, toroidal
orbifolds, such as $T^6/\mbb{Z}_N$ or  $T^6/\mbb{Z}_N\times \mbb{Z}_M$ are
natural candidates for string backgrounds. Therfore, let us discuss
the application of Table \ref{tcs} to such orbifolds in some more detail.

Recall that the spectrum in Table \ref{tcs} is meant to be computed using the
intersection numbers on the resolved orbifold and not on the ambient  torus.
There are some three-cycles $\pi_a$ on the orbifold space  which are inherited
from the torus. In the Kaluza-Klein  reduction on the orbifold they correspond
to massless modes in the untwisted closed string sector of the theory. In general
three-cycles $\pi^t_a$ on the torus are arranged in orbits of length $N$ under
a $\mbb{Z}_N$ orbifold group, i.e., \be \pi^o_a = \sum_{j=0}^{N-1}
\Theta^j\,\pi^t_a , \ee where $\Theta$ denotes the generator of  $\mbb{Z}_N$.
Such an orbit can then be considered as a three-cycle of the orbifold, where
the intersection number is given by \be \pi^o_a\circ\pi^o_b={1\over N}
\left(\sum_{j=0}^{N-1} \Theta^j\, \pi^t_a
  \right) \circ \left(\sum_{k=0}^{N-1} \Theta^k\, \pi^t_b \right) .
\ee
Beside these untwisted three-cycles, certain twisted sectors of the
orbifold action can give rise to additional so-called twisted three-cycles,
which correspond to massless fields in the twisted sectors of the orbifold.
Since these twisted three-cycles are not explicitly needed in this article,
we refer the reader to the existing literature \cite{bbkl02,bgo02,GH03,ho04} to see
how these twisted cycles can be appropriately dealt with.

 Table \ref{tcs} only gives the chiral spectrum
of an intersecting D6-brane model. To compute the generally moduli dependent
non-chiral spectrum one has to employ the usual techniques of conformal field
theory. Therefore, the Higgs sector of a given model is
under less analytic control than the chiral matter sector.

\subsection{Generalized Green-Schwarz Mechanism}
\label{ssggsm} Given the chiral spectrum of Table \ref{tcs}, we have stated
 that the non-Abelian gauge anomalies of all $SU(N_a)$ factors in the gauge group
vanish. On the other hand, the Abelian, the mixed Abelian-non-Abelian and the
mixed Abelian-gravitational anomalies  naively do not. However, as string
theory is a consistent theory, it provides another mechanism to cancel these
anomalies. This is the so-called Green-Schwarz mechanism \cite{gs84}
which can be  generalized to the intersecting D-brane case
\cite{afiru00a}. Here let us discuss  in some more detail the
 mixed Abelian-non-Abelian  anomalies.

\begin{figure}
\begin{center}
\epsfbox{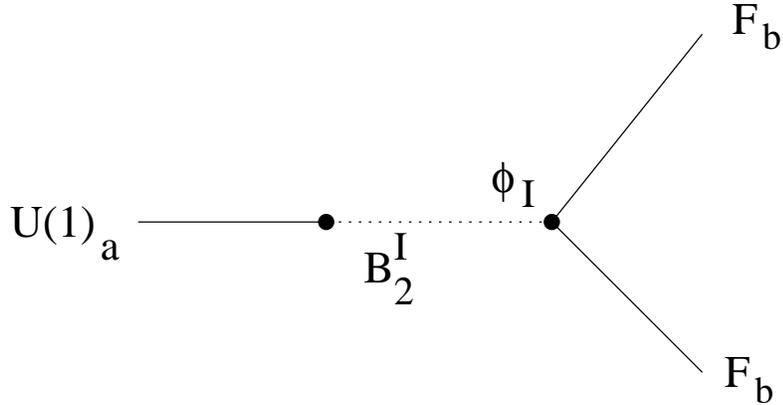} \caption{Green-Schwarz mechanism} \label{fgsm}
\end{center}
\end{figure}

Computing the $U(1)_a-SU(N_b)^2$ anomalies in the effective four-dimensional
gauge theory one finds \be \label{mixedan} A^{}_{ab}={N_a \over 2}\left(
-\pi_a+\pi'_a\right)\circ \pi_b . \ee for each  pair of stacks of D-branes. On
each stack of D6-branes there exist Chern-Simons couplings of the form \be
\label{gsa} \int_{\IR^{1,3}\times\pi_a}  C_3\wedge {\rm Tr}\left(F_a\wedge
F_a\right) , \quad\quad \int_{\IR^{1,3}\times\pi_a}  C_5\wedge {\rm
Tr}\left(F_a\right) \ee where $F_a$ denotes the gauge field on the
D$6_a$-brane. Now we expand every three-cycle $\pi_a$ and $\pi'_a$  into an
integral basis $(\alpha^I,\beta_J)$ of $H_3(M,\mbb{Z})$ with $I,J=0, \ldots,
h_{21}$.
\be \pi_a=e^a_I\, \alpha^I+m^J_a\, \beta_J , \quad\quad
                    \pi'_a=(e^a_I)'\, \alpha^I+(m^J_a)'\, \beta_J .
\ee
This allows us to define  the four-dimensional axions $\Phi_I$ and 2-forms
$B^I$  as
\bea \Phi_I&=&\int_{\alpha^I} C_3,
\quad\quad
                      \Phi^{I+h^{(2,1)}+1}=\int_{\beta_I} C_3, \nonumber \\
                      B^I&=&\int_{\beta_I} C_5, \quad\quad
                      B_{I+h^{(2,1)}+1}=\int_{\alpha^I} C_5 .
\eea In four dimensions $(d\Phi_I, dB^I)$ and
$(d\Phi^{I+h^{(2,1)}+1},dB_{I+h^{(2,1)}+1})$ are Hodge dual to each other. The
general couplings (\ref{gsa}) can now be dimensionally reduced to four
dimensions and yield axionic couplings of the form \bea \int_{\IR^{1,3}}
\Phi_I \wedge {\rm Tr}\left(F_a\wedge
                 F_a\right), \quad\quad  \int_{\IR^{1,3}} B^I \wedge F_a.
\eea The tree-level contribution to the mixed gauge anomaly described by these
couplings takes the form depicted in Figure \ref{fgsm},
\noindent and, adding up all these terms taking the R-R tadpole conditions
into account, one can show that the result has
precisely the form (\ref{mixedan}) and cancels the field theoretic anomaly
\cite{bbkl02}. By the same mechanism also the Abelian and mixed
gravitational-gauge  anomalies are canceled, where the latter ones arise  from
the $U(1)_a-G-G$ triangle diagram and are given by \be \label{gravian}
A^{(G)}_{a}={3\,N_a }\, \pi_{O6} \circ \pi_a. \ee This anomaly is canceled by
the Chern-Simons coupling \be \int_{\IR^{1,3}\times\pi_a}  C_3\wedge {\rm
Tr}\left(R \wedge R\right). \ee
A second important effect of these couplings
is that some of the $U(1)$ gauge fields pair up with the axions to
become a massive gauge field.
The axionic couplings have the detailed form
\be
  \int_{\IR^{1,3}} N_a (\pi_{a,I}-\pi'_{a,I})\, B^I \wedge F_a
\ee
with $\pi_{a,I}\in\{e^a_I,m_a^I\}$ depending on the index $I$.
It has been pointed out in \cite{imr01,LI01} that in
general not only the anomalous $U(1)$s receive a mass, but also some of the
anomaly-free ones, which are given by the kernel of the matrix
\be
  M_a^I  =N_a (\pi_{a,I}-\pi'_{a,I}).
\ee
Therefore,  to determine the low energy spectrum one has to
carefully analyze these quadratic couplings. The massive $U(1)$s still give
rise to  perturbative global $U(1)$ symmetries of the low-energy 
effective action \cite{iq99}, which severely constrain the allowed couplings.

\subsection{Supersymmetric D-branes}
\label{sssb} So far we were not assuming anything more about the D6-branes
than that they are wrapping some homological three-cycles in the background
geometry. If one is interested in supersymmetric models, further constraints
on the bulk geometry
and the cycles on which the D6-branes wrap have to be imposed. Throughout this
section we assume that ${\cal M}$ is a Calabi-Yau manifold so that the closed
string bulk sector  of the Type IIA orientifold preserves ${\cal N}=1$
supersymmetry. First of all, one has to require that each D-brane by itself
preserves supersymmetry, i.e.,  it has to be a BPS brane. As was shown in
\cite{bbs95} this implies that the three-cycles the D6-branes are allowed to
wrap have to be so-called special Lagrangian (sLag) cycles, which are defined
as follows.

On a Calabi-Yau manifold there exist a covariantly constant holomorphic
three-form, $\Omega_3$, and a K\"ahler 2-form $J$. A three-cycle $\pi_a$ is
called Lagrangian if the restriction of the K\"ahler form on the cycle
vanishes \be J\vert_{\pi_a} =0  . \ee If the three-cycle in addition is volume
minimizing, which can be expressed as the property that the imaginary part of
the three-form $\Omega_3$ vanishes when restricted to the cycle, \be
\label{calia} \Im(e^{i\varphi_a}\, \Omega_3)\vert_{\pi_a} =0  ,\quad \ee then
the three-cycle is called a sLag cycle.  The parameter $\varphi_a$ determines
which ${\cal N}=1$ supersymmetry is preserved by the brane. Thus, different
branes with different values for $\varphi_a$ preserve different ${\cal N}=1$
supersymmetries. One can show that (\ref{calia})  implies that the volume of
the three-cycle is given by \be { \label{calib} \rm Vol}(\pi_a)=\left\vert
\int_{\pi_a} \Re(e^{i\varphi_a}\, \Omega_3). \right\vert \ee A shift of
$\varphi_a\to \varphi_a+\pi$ corresponds to exchanging a D-brane by its
anti-D-brane, where the D-brane really satisfies (\ref{calib}) without taking
the absolute value. Therefore a supersymmetric  cycle $\pi_a$ is calibrated
with respect to $\Re(e^{i\varphi_a}\Omega_3)$.

Let us define locally the holomorphic $3$-form $\Omega_3$ and the K\"ahler
form $J$ by \be \label{OJloc} \Omega_3 = dz_1 \wedge dz_2 \wedge dz_3 , \quad
J = i \sum_{i=1}^3 { dz_i \wedge d\bar{z}_i} . \ee Let us choose
$\overline\sigma$ to be just complex conjugation in local coordinates. Then
from $\overline\sigma ( \Omega_3 ) = \overline\Omega_3$ and $\overline\sigma( J ) = -J$ it follows
that the fixed three-cycle of the anti-holomorphic involution is a sLag cycle
with $\varphi_a=0$. Therefore, to finally obtain a globally ${\cal N}=1$
supersymmetric intersecting D-brane model all D6-branes have to wrap sLag
three-cycles which are calibrated with respect to the same three-form
$\Re(\Omega_3)$. It has been checked  in \cite{cim02,bbkl02} that for such globally
supersymmetric configurations indeed the NS-NS tadpoles cancel precisely if
the  R-R tadpoles are canceled.
One can indeed show that precisely if two
branes are relatively supersymmetric one of the four complex world-sheet
bosons becomes massless and extends the massless chiral fermion at the
intersection point to a complete ${\cal N}=1$ chiral supermultiplet.

In  section \ref{sssibt} we shall see that  for the  toroidal (orbifold)
compactifications this  condition becomes a simple geometric condition on the
intersection angles of each D-brane with respect to the orientifold plane.

\subsection{Lift to $G_2$ Compactifications of M Theory}
\label{ssg2}
In this section we briefly discuss the relation between
intersecting D6-brane models and M-theory compactifications on
$G_2$ manifolds. This needs some more advanced mathematical notions
and is not really relevant for understanding the rest of the review.
For completeness, however, we summarize some key ideas here.

Globally ${\cal N}=1$ supersymmetric intersecting D6-brane models
have also shed light on how chiral fermions arise in $G_2$ compactifications
of M theory.
D6-branes and O6-planes are special because they correspond to pure geometry
at strong coupling (unlike other branes which carry additional sources, i.e.,
M-branes or G-fluxes). Therefore, from the number of supercharges the
background preserves, the globally ${\cal N}=1$ supersymmetric intersecting
D6-brane models are expected to lift up to eleven-dimensional M-theory
compactification on singular $G_2$ manifolds
\cite{aw01,csu01a,EW01,acw01,csu01b}. In the Type IIA picture, chiral fermions
are localized at the intersection of D6-branes. Away from the intersections of
IIA D6-branes and/or O6-planes, the IIA configuration corresponds to D6-branes
and O6-planes wrapped on (disjoint) smooth supersymmetric three-cycles, which
we denote generically by $Q$. The corresponding $G_2$  holonomy space hence
corresponds to fibering a suitable Hyperk\" ahler four-manifold over each
component of $Q$. That is an $A$-type ALE singularity for $N$ overlapping
D6-branes, and a $D$-type ALE space for D6-branes on top of O6-planes (with
the Atiyah-Hitchin manifold for no D6-brane, and its double covering for two
D6-branes etc., as follows from \cite{s96,sw96}). Intersections of objects in
type IIA therefore lift to co-dimension 7-singularities, which are isolated up
to orbifold singularities. It is evident from the IIA picture that the chiral
fermions are localized at these singularities.

The structure of these singularities has been studied directly in the $G_2$
context in \cite{aw01}. One starts by considering the (possibly partial)
smoothing of a Hyperk\" ahler ADE singularity to a milder singular space,
parameterized by a triplet of resolution parameters (D-terms or moment maps in
the Hyperk\" ahler construction of the space). The kind of 7-dimensional
singularities of interest are obtained by considering a three-dimensional base
parameterizing the resolution parameters, on which one fibers the
corresponding resolved Hyperk\" ahler space. The geometry is said to be the
unfolding of the higher singularity into the lower one. This construction
guarantees that the total geometry admits a $G_2$ holonomy metric. To
determine the matter content arising from the singularity, one decomposes the
adjoint representation of the A-D-E group associated with the higher
singularity with respect to that of the lower. One obtains chiral fermions
with quantum numbers in the corresponding coset, and multiplicity given by an
index which for an isolated singularity is one.
This construction arises in the M theory lift of the intersecting D6-brane
models. For example, at points where two stacks of $N$ D6-branes and $M$
D6-branes intersect, the  M theory  lift corresponds to a singularity of the
$G_2$ holonomy space that represents the unfolding of an $A_{M+N-1}$
singularity into a 4-manifold with an $A_{M-1}$ and an $A_{N-1}$ singularity.
By the decomposition of the adjoint  representation of $A_{M+N-1}$, we expect
the charged matter to be in the bi-fundamental representation of the
$SU(N)\times SU(M)$ gauge group, in agreement with the IIA picture. A
different kind of intersection arises when $N$ D6-branes intersect with an
O6-plane, and consequently with the $N$ D6-brane images. The M theory lift
corresponds to the unfolding of a $D_N$ type  singularity into an $A_{N-1}$
singularity. The decomposition of the adjoint representation  predicts the
appearance of chiral fermions in the antisymmetric representation of $SU(N)$,
in agreement with the IIA picture.

\subsection{Examples}
\label{sse} So far we have presented the main conceptual ingredients for
constructing intersecting D6-brane models in a fairly general way. In order to
see how this formalism works, let us work out two simple examples in more
detail.

\subsubsection{Intersecting D6-branes on the torus}
\label{sssibt} 
\begin{figure}
\begin{center}
\epsfbox{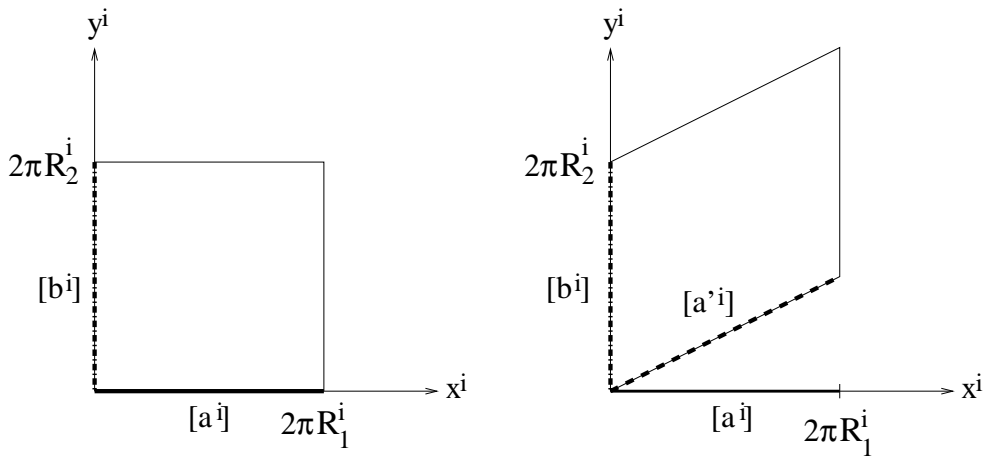} \caption{Choices of $T^2$s} \label{fct2}
\end{center}
\label{choicesofT^2}
\end{figure}

As in section \ref{ssc}  we assume that the six-dimensional
torus factorizes as $T^6=T^2\times T^2\times T^2$. Introducing complex
coordinates $z^i=x^{i}+i y^{i}$ on the three $T^2$ factors, the
anti-holomorphic involution $\overline\sigma$ is chosen to be just complex conjugation $z^i\to
\overline z^i$. Then as shown in Figure \ref{fct2}, on each $T^2$ there exist two
different choices of the complex structure, which are consistent with the
anti-holomorphic involution.
Next we introduce factorizable D6-branes, which are specified by wrapping
numbers $(n^i,m^i)$ along the fundamental cycles $[a^i]$ and $[b^i]$
respectively $[a'^i]$ and $[b^i]$  on each $T^2$. It is useful to express also
the branes for the tilted tori in terms of the untilted 1-cycles $[a^i]$ and
$[b^i]$ by writing $[a'^i]=[a^i]+{1\over 2}[b^i]$. Then a  three-cycle can be
written as a product of three 1-cycles \be \label{torcycle}
\pi_a=\prod_{i=1}^3 \left( n^i_a\ [a^i] + \widetilde m^i_a\ [b^i] \right). \ee
with $\widetilde m^i_a=m^i_a$ for untilted tori  and $\widetilde
m^i_a=m^i_a+{1\over 2}n^i_a$ for  tilted ones. Using the fundamental
intersection number $[a^i]\circ[b^i]=-1$ with all the remaining ones vanishing,
the intersection number between two three-cycles can be computed as \be
\label{inti}
     I_{ab}=\prod_{i=1}^3 (n^i_a\, \widetilde m^i_b- \widetilde m^i_a\, n^i_b)
           =\prod_{i=1}^3 (n^i_a\, m^i_b- m^i_a\, n^i_b).
\ee To work out the tadpole cancellation conditions one has to determine the
three-cycle of the O6-plane and the action of the anti-holomorphic
involution on the  D6-branes.

Independent of the tilt on each $T^2$, the O6-plane is wrapping the cycle
$2[a^i]$, so that the entire three-cycle reads $\pi_{O6}=8\prod_i [a^i]$. The
action of $\overline\sigma$ on a general three-cycle is simply $(n^i,\widetilde
m^i)\to (n^i,-\widetilde m^i)$. Expanding the general tadpole cancellation
condition 
for the homological R-R charges (\ref{tadpole}), one obtains the four independent equations \bea
\label{tad}
\ [a^1][a^2][a^3] &:&    \sum_{a=1}^K   N_a \prod_i n^i_a = 16 \nonumber \\
\ [a^i][b^j][b^k] &:&    \sum_{a=1}^K   N_a\, n^i_a\,  \widetilde m^j_a\,
\widetilde m^k_a =0, \ {\rm with}\ i\ne j\ne k \ne i. \eea 
These formulas were 
 derived initially in \cite{bgkl00a} using a conformal field theory
approach. As discussed in subsection \ref{ssrrtc}, these tadpole conditions
should be supplemented with some additional K-theory constraints.
The K-theory constraints for the toroidal orientifold
were derived in \cite{FM03}, 
and together
with the tadpole conditions (\ref{tad}) above
provide the main constraints  for building  semi-realistic
Standard-like models 
in toroidal orientifolds.

In evaluating the supersymmetry conditions we first consider the non-compact
situation and  a factorizable D-brane which intersects the $x^i$-axes on each
$T^2$ at an  angle $\varphi_a^i$. With $J$ and $\Omega_3$ chosen as in
(\ref{OJloc}), we notice that a factorizable D-brane always satisfies
$J|_{\pi_a}=0$. Expanding the second condition $\Im(\Omega_3)|_{\pi_a}=0$
leads to \be
  0= (dy^1\,dy^2\,dy^3-dy^1\,dx^2\,dx^3  - dx^1\,dy^2\,dx^3 - dx^1\,dx^2\,dy^3)|_{\pi_a}.
\ee
Using  ${dy^i\over dx^i}|_{\pi_a}={\widetilde m^i_a \over n^i_a} u^i$ with
$u^i={R^i_2\over R^i_1}$, this can be brought to the form
\be
\prod_{i=1}^3 \widetilde m^i_a - \sum_{i\ne j\ne k \ne i} \widetilde m^i_a\,
    n^j_a\, n^k_a (u^j\, u^k)^{-1} =0.
\ee
A further constraint arises from the condition $\Re(\Omega_3)|_{\pi_a}>0$,
which takes the form
\be
\prod_{i=1}^3  n^i_a - \sum_{i\ne j\ne k \ne i} n^i_a\,
    \widetilde m^j_a\, \widetilde m^k_a (u^j\, u^k) >0.
\ee
These two conditions are equivalent to the maybe more familiar
supersymmetry condition
\be
\label{susy} \phi_a^1+\phi_a^2+\phi_a^3=0\ {\rm  mod}\ 2\pi.
\ee

We conclude that for a  given D-brane with definite wrapping numbers the
supersymmetry condition (\ref{susy}) puts a constraint on the complex
structure moduli $u^i = {R_2^i\over  R^i_1}$. If all $\phi_a^i\ne 0$ then the
D6-branes preserve ${\cal N}=1$ supersymmetry, and if some angles are
vanishing either  ${\cal N}=2$ or the maximal ${\cal N}=4$ supersymmetry is
preserved.

It has been shown that using factorizable branes on $T^6$ no non-trivial globally
supersymmetric, tadpole cancelling configuration  of intersecting D6-branes
exists. The physical reason for this is that the moment the D6-branes do not
lie entirely on the x-axes, the tension of the branes in the perpendicular
y-directions cannot be compensated, as there are no orientifold planes with
negative tension along these directions. In the T-dual picture with magnetic
fluxes, more general non-factorizable configurations of branes were
investigated \cite{AM04,lmrs05}.

\subsubsection{Intersecting D6-branes on the $\mbb{Z}_2\times \mbb{Z}_2$ orbifold}
\label{IntersectingZ2xZ2}
\label{sssibz2} As pointed out at the end of the last section, to
obtain non-trivial supersymmetric models one needs more orientifold planes
extending  also along  y-directions. The easiest way to obtain   these is by
considering not just tori but toroidal orbifolds, of which the
$\mbb{Z}_2\times \mbb{Z}_2$ orbifold is the simplest \cite{fhs00,csu01,csu01a}. The
orbifold action of the two $\mbb{Z}_2$ symmetries is defined as \be
  \Theta:\cases{ z_1\to -z_1 \cr
                 z_2\to -z_2 \cr
                  z_3\to z_3 \cr }\quad\quad\quad\quad
   \Theta':\cases{ z_1\to z_1 \cr
                 z_2\to -z_2 \cr
                  z_3\to -z_3 \cr }.
\ee As it stands this model is not completely defined, as there are two
possible choices for the signs of the action of $\Theta'$ in the $\Theta$
twisted sector and vice versa. This freedom is called discrete torsion, and here we consider
the model in which one keeps the $(1,1)$ forms in the twisted sectors and
kills  the $(2,1)$ forms. Therefore, this model has the Hodge numbers
$(h_{21},h_{11})=(3,51)$, which means that there are precisely eight
three-cycles in the untwisted sector. These are \be [a^1][a^2][a^3]^t,\
[a^i][a^j][b^k]^t,\ [a^i][b^j][b^k]^t,\ [b^1][b^2][b^3]^t \ee with $i\ne j\ne
k\ne i$, and where the upper index indicates that so far these three-cycles
are defined on the ambient $T^6$.

Given the rules from section \ref{ssms}  of how to deal with three-cycles in
the orbifold case, we have to carefully distinguish between three-cycles in
the ambient $T^6$ and three-cycles on the orbifold space. Under the action of
$\mbb{Z}_2\times \mbb{Z}_2$ a three-cycle on $T^6$ has 3 images, which
homologically are identical to the original cycles. Therefore a three-cycle in
the bulk of the orbifold space can be identified with $\pi^B=4 \pi^t$.
Applying the rule for the intersection number we get $\pi_a^B\circ \pi_b^B =4
\pi_a^t\circ \pi_b^t$. Therefore, the cycles $\pi_a^B$ do not span the
integral homology lattice $H_3(M,\mbb{Z})$, which suggests that there exist
smaller three-cycles in the orbifold space. This is indeed the case. By
choosing the three-cycles to run through the origin, we obtain three-cycles
which are given by $\pi^o_a={1\over 2}\pi^B_a$, which have intersections on
the orbifold $\pi^o_a\circ \pi^o_b =\pi_a^t\circ \pi_b^t$. Therefore, the
untwisted three-cycles on the orbifold space have the same form as in
(\ref{torcycle}) and the same intersection form (\ref{inti}) with the only
difference that the basis of three-cycles is now defined on the orbifold space
instead of the torus.

Working out the fixed point locus of the four orientifold projections $\Omega
\overline\sigma(-1)^{F_L}$, $\Omega \overline\sigma\Theta (-1)^{F_L}$,
$\Omega \overline\sigma\Theta'(-1)^{F_L}$,
$\Omega  \overline\sigma\Theta\Theta'(-1)^{F_L}$ and
expressing everything in terms of three-cycles in the orbifold, we obtain \be
\pi_{O6}=4 \prod_i [a^i]^o -\sum_{i\ne j\ne k\ne i}
     4^{1-\beta_j-\beta_k} [a^i][b^j][b^k]^o,
\ee where $\beta_j=0$ for an untilted $T^2$ factor and $\beta_j=1/2$ for a
tilted one. Therefore, the four tadpole cancellation conditions 
for the homological R-R charges
now read
\footnote{In \cite{csu01b} a different  convention was used such that
there appeared an overall factor of two in all four
R-R tadpole cancellation in conditions (\ref{tadorb}).
This is consistent with the rank of the gauge group, as the
rule in  \cite{csu01b} was that a stack of $N_a$ branes
carries a gauge group $U(N_a/2)$.}
\bea
\label{tadorb}
\ [a^1][a^2][a^3]^o &:&    \sum_{a=1}^K   N_a \prod_i n^i_a = 8, \nonumber \\
\ [a^i][b^j][b^k]^o &:&    \sum_{a=1}^K   N_a\, n^i_a\,  \widetilde m^j_a\,
\widetilde m^k_a =
   -2^{3-2\beta_j-2\beta_k},\     {\rm with}\ i\ne j\ne k \ne i.
\eea Note the changes  on the right hand side of (\ref{tadorb}) as compared to
the purely toroidal case (\ref{tad}). 
To ensure consistency of the models,
these tadpole conditions should be supplemented with additional K-theory
constraints. For untilted tori, 
the K-theory constraints for this $\mbb{Z}_2 \times \mbb{Z}_2$ orientifold 
\cite{ms04a} read:
\bea
\label{tadpolesK}
\sum_{a=1}^K   N_a \prod_i m^i_a & \in &  2 \mbb{Z}, \nonumber \\
\sum_{a=1}^K   N_a\, n^i_a\,  n^j_a\, m^k_a & \in & 2 \mbb{Z},\     {\rm with}\ i\ne j\ne k \ne i.
\eea
It is straightforward to generalize these conditions to cases where some or allof the tori
are tilted.
 
Finally, for the intersection number
between a D-brane and the orientifold plane one obtains \be \pi_{O6}\circ
\pi^o_a=4 \prod_i \widetilde m^i_a - \sum_{i\ne j\ne k\ne i}
            4^{1-\beta_j-\beta_k}\, \widetilde m^i_a\, n^j_a\, n^k_a .
\ee The supersymmetry conditions are the same as for the toroidal  case.

The equations developed in the last two subsections provide   the main tools
for constructing quasi-realistic intersecting D-brane models in these two most
simple backgrounds.

\section{SEMI-REALISTIC INTERSECTING \\ D-BRANE MODELS}
\label{ssrm} In this section we give an overview of the different intersecting
D-brane world models explicitly constructed so far. Essentially, there are two
philosophical attitudes towards approaching this problem, which differ in
their assumptions about the size of the string scale, i.e., the energy scale
where stringy effects become relevant. In particular, because of stability and
phenomenological considerations it is usually assumed that $M_s$ is low (e.g.,
$1-100$ TeV) for non-supersymmetric constructions, while supersymmetric
studies usually assumed that $M_s$ is much closer to the Planck scale.

\subsection{Non-supersymmetric Standard-like Models}
\label{ssnssm} Here we review different approaches to construct semi-realistic
non-supersymme\-tric standard-like models and highlight some fairly general
phenomenological features of such models. The first explicit chiral
intersecting D-brane  models were constructed in
\cite{bgkl00a,aads00,afiru00,afiru00a,bkl00,imr01}, 
where, except in \cite{aads00,afiru00,afiru00a},
the background space was simply chosen to be an $\Omega \overline\sigma
(-1)^{F_L}$ orientifold of a factorisable torus $T^6$. These articles
triggered a lot of subsequent work using essentially the same framework and
ideas but generalizing the D-brane set-ups in certain ways. Before we list all
these different constructions, we would like to present a proto-type model,
which shows that the particle content of intersecting D-brane models can come
quite close to the Standard Model.

\subsubsection{A simple semi-realistic model}

\label{sssssrm} In \cite{imr01} the authors were considering the simple
toroidal orientifold set-up mentioned above. Using a bottom-up approach, they
introduced four stacks of D6-branes with the wrapping numbers chosen as shown
in Table \ref{twrap}.
\begin{table}
\caption{Wrapping numbers for a semi-realistic non-supersymmetric model.
The parameters are defined
as $\overline\beta^{1,2}=1-\beta^{1,2}$, $\beta^3=1/2$, $\rho=1,1/3$, $\epsilon=\pm
1$ and $n^2_a,n^1_b,n^1_c,n_d^2\in \mbb{Z}$.} \centering \vspace{3mm}
\label{twrap}
\begin{tabular}{|c||c|c|c|}
\hline
$N_a$ & $(n^1,\widetilde m^1)$ & $(n^2,\widetilde m^2)$ & $(n^3,\widetilde m^3)$  \\
\hline\hline
$ N_a=3$ &  $(1/\overline\beta^1,0)$ & $(n^2_a,-\epsilon \overline\beta^2)$ & $(1/\rho,-1/2)$ \\
\hline
$ N_b=2$ &  $(n^1_b,\epsilon \overline\beta^1)$ &  $(1/\overline\beta^2,0)$ & $(1,-3\rho/2)$ \\
\hline
$ N_c=1$ &  $(n^1_c,-3\rho\epsilon \overline\beta^1)$ & $(1/\overline\beta^2,0)$ & $(0,-1)$ \\
\hline
$ N_d=1$ &  $(1/\overline\beta^1,0)$ & $(n^2_d,\overline\beta^2\epsilon/\rho)$ & $(1,-3\rho/2)$ \\
\hline
\end{tabular}
\end{table}
The intersection numbers between these four stacks of D6-branes give rise to
the chiral fermions listed in Table \ref{tferm}, which transform in the
various bi-fundamental representations of the (naive) gauge group
$SU(3)_C\times SU(2)_W\times U(1)_a\times U(1)_b\times U(1)_c\times U(1)_d$.
\begin{table}
\caption{Chiral massless spectrum of the semi-realistic four stack model.
$(.)^c$ denotes the charge conjugated field.}
\centering \vspace{3mm} \label{tferm}
\begin{tabular}{|c|c|c|c|}
\hline
Intersection & Matter & Rep. & $Y$  \\
\hline\hline
$(a,b)$ & $Q_L$ & $(3,2)_{(1,-1,0,0)}$ \hfill & $1/6$ \\
$(a',b)$ & $q_L$ & $2\times (3,2)_{(1,1,0,0)}$ & $1/6$ \\
$(a,c)$ & $(U_R)^c$ & $3\times (\overline 3,1)_{(-1,0,1,0)}$ & $-2/3$ \\
$(a',c)$ & $(D_R)^c$ & $3\times (\overline 3,1)_{(-1,0,-1,0)}$ & $1/3$ \\
$(b',d)$ & $L_L$ & $3\times (1,2)_{(0,-1,0,-1)}$ & $-1/2$ \\
$(c,d)$ & $(E_R)^c$ & $3\times (1,1)_{(0,0,-1,1)}$ & $1$ \\
$(c',d)$ & $(N_R)^c$ & $3\times (1,1)_{(0,0,1,1)}$ & $0$ \\
\hline
\end{tabular}
\end{table}
The hypercharge is given by the linear combination $Q_Y={1\over 6} Q_a
-{1\over 2}Q_c +{1\over 2}Q_d$. For more details and the phenomenological
implications we refer the reader to \cite{imr01}. Here we would like to simply
list some 
features typical for such intersecting D-brane
models:

\begin{itemize}

\item{There are many (infinite)  non-supersymmetric intersecting D-brane
constructions  with
      the Standard Model particle spectrum, where in most cases one needs
      additional ``hidden'' branes to satisfy tadpole cancellation.}

\item{Models with only bi-fundamental matter necessarily contain right
     handed neutrinos. }

\item{One obtains additional $U(1)$ factors, which partly receive  a mass via
    the  generalized Green-Schwarz mechanism (see section \ref{ssggsm}); the
condition
     that $U(1)_Y$ remains massless imposes further conditions on the parameters
     in Table \ref{twrap}. }

\item{All the massive former  $U(1)$ gauge symmetries survive as perturbative
   global symmetries
  and can be identified with baryon number $Q_a$ and lepton
   number $Q_d$, stabilizing the proton and preventing Majorana neutrino masses.}

\item{One can show that in the $(bc)$ sector there are additional non-chiral
(tachyonic)  fields, which might have an interpretation as Higgs
particles; condensation of these fields corresponds to D-brane        
recombination in string theory (see for instance \cite{ht97,afiru00a,cim02a,hn03,eghk03,el03}).}

\end{itemize}
Since the models are not supersymmetric, there are  typically uncanceled
NS-NS tadpoles, contributing to the (dilaton dependent)
cosmological constant which is of the order of $M_s^4$. In addition, in the
effective theory below the string scale there are large radiative corrections
of the order of  $M_s$. Therefore,  typically these models require $M_s$ of
the order of  the TeV scale. However, as emphasized in the subsection
\ref{ssss}, for the toroidal  constructions with intersecting
D6-branes  the internal space cannot be much larger than the Planck volume,
and  the string scale $M_s$ is restricted to be of the order of
the Planck scale.\footnote{This is not, however, a fundamental problem of the intersecting brane 
world scenario, since the chiral spectrum of Table \ref{tferm} can be achieved in 
models where the string scale can be lowered to a TeV \cite{cim02b}.}

\subsubsection{Generalizations}

\label{sssg} Many generalizations of the above construction have been
considered in the literature. Here we only list, in non-chronological order,
the ones for which only non-supersymmetric models are possible or have been
considered. For more details on the various constructions we refer the reader
to the original literature.

The straightforward generalization of the above set-up is to introduce more
than 4 four stacks of D6-branes \cite{CK02a,CK02b,CK02e} to realize directly
the Standard Model gauge group. Similarly, one can try to find 
Grand-Unified-like models
in this toroidal set-up \cite{CK02,CK02d}.

If one is giving up supersymmetry, then of course there is no need to
introduce orientifold planes in the first place, and one can simply start
with intersecting D6-branes in Type IIA \cite{afiru00,afiru00a}.

Another approach is not to work with D6-branes but instead with D4-
respectively D5-branes, where in order to achieve chirality one has to
perform an additional orbifold in the transverse space \cite{afiru00,afiru00a}. 
Therefore, the models constructed in
\cite{afiru00,afiru00a,fhs01,bkl01,GH01,GH02,cim02b,CK02c,bkl02a,bkl02b,DB02,bkl02c}
can be regarded as a hybrid of the two ways to obtain chiral fermions, namely
as intersecting branes at singularities.

Giving up supersymmetry one can also start with orientifolds of Type O string
theory \cite{bkl02}.

A peculiarity about intersecting D-branes has been pointed out in
\cite{cim02,cim02a,cim02c}, namely that one can build models in which at each
intersection between two branes an ${\cal N}=1$ supersymmetry is preserved,
even though it is not preserved globally. In such models  the absence of
one-loop corrections to the Higgs mass weakens the gauge hierarchy problem and
allows one to enhance the string scale up to 10 TeV.
Such so-called quasi-supersymmetric models have also been 
studied in \cite{MK02,MK02a} from a field theory perspective, 
and additional models have been constructed in \cite{ll03}.

In \cite{bklo01,bklo01a} Type IIA orientifolds on the $\mbb{Z}_3$ orbifold
were considered and a non-supersymmetric three-generation flipped $SU(5)$ 
Grand Unified Model \cite{ekn02} was constructed explicitly.

\subsection{Supersymmetric models  on the $\mbb{Z}_2\times \mbb{Z}_2$ orientifold}
\label{sssz2} In order to show how semi-realistic supersymmetric
intersecting D-brane models can arise and what their salient features are,
we now describe in some more detail the  four-dimensional
chiral ${\cal N}=1$ supersymmetric
intersecting D-brane models constructed in \cite{csu01,csu01a}. Unlike the
non-supersymmetric models discussed so far, these supersymmetric intersecting
D-brane models are stable because both the NS-NS and the R-R tadpoles are
canceled. As discussed in section \ref{ssg2},
these models also have the additional interesting feature that when
lifted to M theory they correspond to chiral $G_2$ compactifications
\cite{csu01a,csu01b}.

The background geometry of this class of models is the $T^6/{\mbb{Z}_2 \times
\mbb{Z}_2}$ orbifold as described in subsection \ref{sssibz2}.
As explained,
there exist two choices of complex structure of $T^2$ that are
compatible with the orientifold symmetry: rectangular or tilted (see Figure
\ref{fct2}). 
If the Standard Model sector D-branes are not on top of the orientifold
planes and the $T^2$ are rectangular,
as in the toroidal models discussed in \cite{bgkl00a}, the number of
chiral families is even.\footnote{The weak sector can come from
D-branes on top of an orientifold plane since $Sp(2) \simeq SU(2)$,
in which case odd number of families can be obtained without tilted
tori \cite{ms04,ms04a}.}
Hence, we consider models with one tilted $T^2$.
This mildly modifies the closed string sector, but has an important impact on
the open string sector since the number of chiral families 
can now be odd. Due to the smaller
number of O6-planes in tilted configurations, the R-R tadpole conditions
(\ref{tadorb}) are
very stringent for more than one tilted $T^2$, so we focus on models with only
one tilted $T^2$.

To simplify the supersymmetry conditions within our search for realistic
models, we consider a particular Ansatz for the intersection
angles of the branes with the x-axes: $(\phi_1,\phi_2,0)$,
$(\phi_1,0,\phi_3)$ or $(0,\phi_2,\phi_3)$ with $\sum_i \phi_i=0$ for each brane.
Focusing on tilting just
the third torus, the search for theories with $U(3)$ and $U(2)$ gauge factors
carried by branes at angles and three left-handed quarks turns out to be very
constraining, at least within our Ansatz. A D6-brane configuration with
wrapping numbers $(n_a^i,\widetilde{m}_a^i)$ which gives rise to a
three-family supersymmetric Standard-like model is presented in Table \ref{tc3f}.

\begin{table}
[htb] \footnotesize
\renewcommand{\arraystretch}{1.25}
\begin{center}
\caption{D6-brane configuration for the three-family $\mbb{Z}_2 \times
\mbb{Z}_2$ orientifold model.} \vspace{3mm} \label{tc3f}
\begin{tabular}{|c||c|l|l|}
\hline Type & $N_a$ & $(n_a^1,m_a^1) \times
(n_a^2,m_a^2) \times (n_a^3,\widetilde{m}_a^3)$ &  Gauge Group \\
\hline
$A_1$ & 4 & $(0,1)\times(0,-1)\times (2,{\widetilde 0})$ &  $Q_{8}, Q_{8'}$ \\
$A_2$ & 1 & $(1,0) \times(1,0) \times (2,{\widetilde 0})$ &  $Sp(2)_A$ \\
\hline
$B_1$ & 2 & $(1,0) \times (1,-1) \times (1,{\widetilde {3/2}})$ & $SU(2), Q_2$  \\
$B_2$ & 1 & $(1,0) \times (0,1) \times (0,{\widetilde {-1}})$ & $Sp(2)_B$ \\
\hline $C_1$ & 3+1 & $(1,-1) \times (1,0) \times (1,{\widetilde{1/2}})$ &
 $ SU(3), Q_3,Q_1 $ \\
$C_2$ & 2 & $(0,1) \times (1,0) \times (0,{\widetilde{-1}})$  & $Sp(4)$  \\
\hline
\end{tabular}
\end{center}
\end{table}

The four  D6-branes labeled $C_1$ are split into two parallel but not overlapping
stacks of three  and  one branes leading to an adjoint  breaking of $U(4)$ into
 $U(3)\times U(1)$. Consequently, a linear combination of the two $U(1)$'s is actually a
generator within the non-Abelian $SU(4)$ arising for coincident branes. This ensures that
this $U(1)$ is automatically non-anomalous and massless (free of linear
couplings to untwisted moduli) \cite{afiru00a,afiru00,imr01}, which turns out to
be crucial for  the appearance of the Standard Model hypercharge.

For convenience we consider the four  D6-branes labelled $A_1$ to be away from
the O6-planes in all three complex planes. This implies
This leads to two  D6-branes that
can move independently  giving rise to a gauge group $U(1)^2$, plus their
$\Theta$, $\Theta'$ and $\Omega \overline\sigma(-1)^{F_L}$ images.
These $U(1)$'s are also
automatically non-anomalous and massless. In the effective theory, this
corresponds to Higgsing of $USp(8)$ down to $U(1)^2$.

The surviving non-Abelian gauge group is $SU(3)_C\times SU(2)_W\times
Sp(2)\times Sp(2)\times Sp(4)$. The $SU(3)_C\times SU(2)_W$ corresponds to the
MSSM, while the last three factors form a quasi-hidden sector, i.e., most
states are charged under one sector or the other, but there are a few which
couple to both. In addition, there are three non-anomalous $U(1)$ factors and
two anomalous ones. The generators $Q_3$, $Q_1$ and $Q_2$ refer to the $U(1)$
factor within the corresponding $U(n)$, while $Q_8$, $Q_8'$ are the $U(1)$'s
arising from the higgsed $USp(8)$. $Q_3/3$ and $Q_1$ are essentially baryon ($B$) and
lepton ($L$) number, respectively, while $(Q_8+Q_8')/2$ is analogous to the
generator $T_{3R}$ occurring in left-right symmetric extensions of the
Standard Model. The hypercharge is defined as:
\begin{eqnarray}
Q_Y & = & \frac 16 Q_3 - \frac 12 Q_1 + \frac 12 (Q_8+Q_8'). \label{hyper}
\end{eqnarray}
 From the above comments, $Q_Y$  is non-anomalous guaranteeing
that $U(1)_Y$ remains  massless.
There are two additional surviving non-anomalous $U(1)$'s, i.e.,
$B-L = Q_3/3-Q_1$ and $Q_8-Q_8'$. The gauge bosons corresponding to the
anomalous $U(1)$ generators $B+L$ and $Q_2$ acquire string-scale masses, so
those generators act like perturbative
global symmetries on the effective four-dimensional
theory.

The spectrum of chiral multiplets in the open string sector is tabulated in
Table \ref{ts3f}.
There are also vector-like multiplets in the model but they are generically
massive so we do not tabulate them here (they can be found in \cite{cls02}).
The theory contains
three Standard Model families, multiple Higgs candidates, a number of exotic
chiral (but anomaly-free)  fields, and multiplets which transform in the
adjoint or singlet representation   of the Standard Model gauge group.

\begin{table} \footnotesize
\renewcommand{\arraystretch}{1.25}
\begin{center}
\begin{tabular}{|c||c||c||c|c|c|}
\hline Sector & $SU(3)_C\times SU(2)_Y\times Sp(2)_B\times Sp(2)_A\times
Sp(4)$ & $\left( Q_3,Q_1,Q_2,Q_8,Q_8' \right)$ & $Q_Y$ & $Q_8-Q_8'$ & Field
\\
\hline $A_1 B_1$ & $3 \times 2\times (1,{\overline 2},1,1,1)$ &
$(0,0,-1,\pm 1,0)$ & $\pm \frac 12$ & $\pm 1$ & $H_U$, $H_D$\\
          & $3\times 2\times (1,{\overline 2},1,1,1)$ &
$(0,0,-1,0,\pm 1)$ & $\pm \frac 12$ & $\mp 1$ & $H_U$, $H_D$\\
$A_1 C_1$ & $2 \times (\overline{3},1,1,1,1)$ & $(-1,0,0,\pm 1,0)$ & $\frac
13, -\frac 23$ & $1,-1$ &
$\bar{D}$, $\bar{U}$\\
          & $2 \times (\overline{3},1,1,1,1)$ &
$(-1,0,0,0,\pm 1)$ & $\frac 13, -\frac 23$ & $-1,1$ &
$\bar{D}$, $\bar{U}$\\
          & $2 \times (1,1,1,1,1)$ &
$(0,-1,0,\pm 1,0)$ & $1,0$ & $1,-1$ &
$\bar{E}$, $\bar{N}$\\
          & $2 \times (1,1,1,1,1)$ &
$(0,-1,0,0,\pm 1)$ & $1,0$ & $-1,1$ &
$\bar{E}$, $\bar{N}$\\
$B_1 C_1$ & $(3,{\overline 2},1,1,1)$ &
$(1,0,-1,0,0)$ & $\frac 16$ & 0 & $Q_L$\\
             & $(1,{\overline 2},1,1,1)$ &
$(0,1,-1,0,0)$ & $-\frac 12$ & 0 & $L$\\
$B_1 C_2$ & $(1,2,1,1,4)$ &
$(0,0 ,1,0,0)$ & 0 & 0 & \\
$B_2 C_1$ & $(3,1,2,1,1)$ &
$(1,0,0,0,0)$ & $\frac 16$ & 0 & \\
          & $(1,1,2,1,1)$ &
$(0,1,0,0,0)$ & $-\frac 12$ & 0 & \\
$B_1 C_1^{\prime}$ & $2\times (3,2,1,1,1)$ &
$(1,0,1,0,0)$ & $\frac 16$ & 0 & $Q_L$ \\
                   & $2\times (1,2,1,1,1)$ &
$(0,1,1,0,0)$ & $-\frac 12$ & 0 & $L$ \\
\hline $B_1 B_1^{\prime}$ & $2\times (1,1,1,1,1)$ &
$(0,0,-2,0,0)$ & 0 & 0 &  \\
                   & $2\times (1,3,1,1,1)$ &
$(0,0,2,0,0)$ & 0 & 0 & \\
\hline \hline
$A_1 A_1$ & $3 \times 8 \times (1,1,1,1,1)$ & $(0,0,0,0,0)$ & 0 & 0 &  \\
& $3 \times 4 \times (1,1,1,1,1)$ & $(0,0,0,\pm 1,\pm 1)$ & $\pm 1$
& 0 &  \\
& $3 \times 4 \times (1,1,1,1,1)$ & $(0,0,0,\pm 1,\mp 1)$ & 0
& $\pm 2$ &  \\
& $3 \times  (1,1,1,1,1)$ & $(0,0,0,\pm 2,0)$ & $\pm 1$ & $\pm 2$ &  \\
& $3 \times  (1,1,1,1,1)$ & $(0,0,0,0,\pm 2)$ & $\pm 1$ & $\mp 2$ &  \\
$A_2 A_2$ & $3 \times (1,1,1,1,1)$ & $(0,0,0,0,0)$ & 0 & 0 &  \\
$B_1 B_1$ & $3 \times (1,3,1,1,1)$ & $(0,0,0,0,0)$ & 0 & 0 &  \\
 & $3 \times (1,1,1,1,1)$ & $(0,0,0,0,0)$ & 0 & 0 &  \\
$B_2 B_2$ & $3 \times (1,1,1,1,1)$ & $(0,0,0,0,0)$ & 0 & 0 &  \\
$C_1 C_1$ & $3 \times (8,1,1,1,1)$ & $(0,0,0,0,0)$ & $0$ & 0 &  \\
& $3 \times (1,1,1,1,1)$ & $(0,0,0,0,0)$ & $0$ & 0 &  \\
$C_2 C_2$ & $3 \times (1,1,1,1,5+1)$ & $(0,0,0,0,0)$ & 0 & 0 &  \\
\hline
\end{tabular}
\end{center}
\caption{\small The chiral spectrum of the open string sector in the
supersymmetric three-family model.
To be complete, we also list  in the bottom part of the
table, below the double horizontal line,  the non-chiral massless states from the $aa$
sectors, which are not localized at the intersections and correspond
to deformation and Wilson line moduli of the $D6$ branes.  } \label{ts3f}
\end{table}

For more details and phenomenological features, please consult
the original literature  \cite{cls02,cls02a}. Here,
we would like to highlight  some of the special features of this
supersymmetric model:
\begin{itemize}
\item The model involves an extended gauge structure, including two additional
$U(1)'$ factors, one of which has family non-universal and therefore flavor
changing couplings. Extended gauge structure is quite generic among string
models, and more so for intersecting D-brane models.

\item There are additional Higgs doublets, suggesting such effects as a rich
spectrum of Higgs particles, neutralinos, and charginos, perhaps with
nonstandard couplings due to mixing and flavor changing effects.

\item In addition to the three chiral families of the Standard Model, there
are chiral exotic states, i.e., chiral states with unconventional Standard Model quantum
numbers. It was argued in \cite{cls02} that these states may decouple from
the low energy spectrum due to hidden sector charge confinement.

\item There exist  a quasi-hidden non-Abelian sector, which becomes strongly
coupled above the electroweak scale. The dynamics of the strongly coupled
hidden sector  leads to dynamical supersymmetry breaking with  dilaton and
untwisted complex structure moduli stabilization, as studied in detail in
\cite{clw03}.
Charge confinement   modifies the low energy  spectrum by causing some exotics
to disappear, while anomaly considerations imply that new composite states may
emerge \cite{cls02} (see also \cite{NK04,NK04a}).

\end{itemize}
Just as in the non-supersymmetric constructions  of  Standard-like models
the model does not have the conventional form of gauge unification, as
each gauge factor is associated with a different set of branes. However, the
string-scale couplings are predicted in terms of the ratio of the Planck and
string scales and a geometric factor as discussed in section \ref{ssss}. The
explicit dependence of the tree-level holomorphic gauge kinetic function
on the dilaton and the
complex structure moduli will be discussed in section \ref{ssgc}. As common to all  intersecting D6-brane constructions, the Yukawa  couplings among
chiral  matter  are due to world-sheet  instantons associated with the string
world-sheet stretching among the  intersections where the corresponding chiral
matter fields are localized \cite{afiru00}. The details of the Yukawa coupling interactions
will be discussed in  Subsection \ref{sssyc}.

\subsubsection{Supersymmetric grand unified models}
\label{ssssgum}

The setup with intersecting D6-branes on orientifolds  also allows for the
construction  of   Grand  Unified Models, based on the Georgi-Glashow $SU(5)$
gauge group \cite{gg74}.  Such non-supersymmetric Grand Unified Models  were
constructed   in \cite{bklo01,ekn02, CK04, CK04b}
and supersymmetric ones in \cite{csu01a,cps02,ckmnw05}.
(For additional work on non-supersymmetric Grand Unified Models see also
\cite{CK02d,CK03}.)

The supersymmetric constructions of  such models  have a  lift  on a circle to
M-theory and provide examples of  Grand Unified  Models of strongly-coupled
M-theory compactified on singular seven  dimensional manifolds with $G_2$
holonomy \cite{aw01,csu01a,EW01,acw01,csu01b}. (See also section \ref{ssg2}.)

The key point in these constructions is  the appearance  of anti-symmetric
representations, i.e., ${\bf 10}$ of $SU(5)$,  which can emerge  at the
intersection of the  D-brane with its orientifold image (see Table \ref{tcs}).
Thus,   ${\bf 10}$-plets, along with the  bi-fundamental representations
$({\overline {\bf 5}}, \, { \bf N_b})$ at the  intersections of $U(5)$ branes
with $U(N_b)$  branes,  form  the chiral particle content of the   quark and
lepton families. It turns out that the gauge boson for the diagonal $U(1)$
factor of $U(5)$ is massive, and the anomalies associated with  this $U(1)$
are canceled via the generalized Green-Schwarz mechanism, as explained in
section \ref{ssggsm}.

For toroidal and ${\mbb{Z}_2 \times \mbb{Z}_2}$ orbifold  compactifications
there are  three copies of the adjoint
representations on the world-volume of the branes. They  are moduli associated
with  the splitting and Wilson lines of the branes that wrap the same
three-cycles which in these two cases are not rigid. Turning  on appropriate vacuum
expectation  values (VEVs)
 of these adjoint representations  can spontaneously  break  $SU(5)$
down to the Standard Model gauge group.
As mentioned such VEVs have a geometric interpretation in
terms of the  appropriate parallel splitting of the $U(5)$ branes.

Since all the Standard Model gauge group factors arise from branes wrapped on
parallel, but otherwise identical cycles, this  construction provides a
natural framework for gauge coupling unification  and thus a natural embedding
of the  traditional grand unification \cite{gg74}  into intersecting D-brane
models.

The  explicit supersymmetric constructions of Grand Unified Models were given
for  the ${\mbb{Z}}_2\times {\mbb{Z}}_2$ orbifold models. The first such
example \cite{csu01a} was a four-family model. Further systematic analysis
\cite{cps02} revealed that within  ${\mbb{Z}}_2\times {\mbb{Z}}_2$ orbifold
models with factorizable three-cycles all the three family models  necessarily
also contain three copies of {\bf 15}-plets, the  symmetric  representations
of $SU(5)$. (Analogous observations have been reported
in \cite{GH04} for supersymmetric $SU(5)$ models in the
${\mbb{Z}}_4\times {\mbb{Z}}_2$ orbifold background.)
 There are approximately twenty such models \cite{cps02}, which
are not fully realistic:
\begin{itemize}
\item The additional {\bf 15}-plets decompose under $SU(3)_C\times SU(2)_Y
\times U(1)_Y$ as as $({\bf 6},{\bf 1})(-\frac{2}{3}) + ({\bf 1},{\bf 3})(+1)
+ ({\bf 3},{\bf 2})(+\frac{1}{6}) $ and thus contain additional exotic
Standard Model particles.

\item Since the chiral states are  charged under the $U(1)$ factor of $U(5)$
and the only candidates for the Higgs fields are in the  adjoint {\bf 24} and
fundamental {\bf 5} representations,   the fermion masses  can arise  only
from the Yukawa couplings of the type: $\overline{\bf 5}$~{\bf 10}
$\overline{\bf 5}_H$ (subscript  $H$ refers to the Higgs fields), while the
couplings of the type ${\bf 10}~{\bf 10}~{\bf 5}_H$ are  absent due to the
$U(1)$ charge conservation \cite{bklo01}. The absence of  perturbative Yukawa couplings
to the up-quark families is generic for these constructions (supersymmetric or  not).

\item Within this framework one can address  the  long standing problem of
doublet-triplet splitting, i.e., ensuring that after the  breaking of $SU(5)$
the doublet of  ${\bf 5}_H$, responsible for the electroweak symmetry
breaking,  remains light  while the triplet becomes heavy. The mechanism,
suggested within M-theory on $G_2$ holonomy  manifolds \cite{EW02}, allows for
the $SU(5)$ breaking via Wilson lines with different discrete quantum numbers
for the doublet and the triplet, which in turn forbids the mass term for the
doublet.  However, the Wilson lines in the present context are continuous
rather than discrete, due to the generic non-rigid nature of three cycles on
orbifold compactifications.
Although the current constructions
of the models are not fully realistic,  generalizations to examples with rigid
three-cycles may provide  an avenue to address the appearance of genuinely
discrete Wilson lines.

\end{itemize}

\subsubsection{Systematic search for supersymmetric Standard-like models}
\label{sssss}

In the previous two subsections, we have seen that supersymmetric
Standard-like and Grand Unified Models can be constructed within the 
intersecting
D6-brane framework. Subsequently, systematic searches for  supersymmetric
three family Standard-like models have  been carried out  
within $\mbb{Z}_2\times
\mbb{Z}_2$ orbifold constructions with factorizable three-cycles.

As mentioned above, within this framework the systematic search for
three-family $SU(5)$ Grand Unified Models  \cite{cps02} produced three family
models
 which however necessarily  also contain three copies of {\bf 15}-plets.
However, this feature is  specific to this specific orbifold and it remains to be seen
whether it persists for more general models.

As for the Standard-like model constructions with  gauge group factors arising
from different  intersecting D6-branes,  sets of models  with fewer Higgs
doublets  \cite{cp03a}  were obtained. However, all these three-family models
still possess additional exotics. Subsequently, a systematic search for
supersymmetric
Pati-Salam models based on the left-right  symmetric gauge symmetry $SU(4)_C\times
SU(2)_L \times SU(2)_R$ was presented in \cite{cll04}.
 The  gauge symmetry can be broken down to the
Standard Model one   via D6-brane splitting and a further D- and F-flatness
preserving Higgs mechanism from massless open string states in an ${\cal N}=2$
subsector. Among the models that also possess at least two confining hidden
gauge sectors, where gaugino condensation can in turn trigger supersymmetry
breaking and (some) moduli stabilization, the search revealed eleven models.
Two  models realize
gauge coupling unification of $SU(2)_L$ and $SU(2)_R$ at the string scale. However,
all these models  still possess additional exotic matter.

In another related work \cite{clll04}, the study of  splitting of D6-branes
parallel  to   orientifold planes, within $\mbb{Z}_2\times \mbb{Z}_2$
orientifolds, led to the examples of four-family standard-like orientifold
models  without chiral exotics. The starting point  is a one-family
$U(4)\times Sp(2f)_L\times Sp(2f)_R$, ($f=4$) model which is broken down to a
four-family $U(4)\times U(2)_L\times U(2)_R$ model by parallel splitting of the
D-branes, originally positioned on the O-planes.
The chirality of the model is changed due to the fact that the original branes
were positioned on top of an orientifold {\it singularity}. Both the string
theory and field theory aspects of these specific  D-brane splittings are
discussed in detail in \cite{clll04}.
 
These systematic searches for realistic models seem to suggest that an extended Higgs
sector is ubiquitious in intersecting D-brane models. It is therefore quite remarkable that a 
simple D-brane configuration, introduced in \cite{cim02d}, yields just the MSSM chiral 
spectrum and its minimal Higgs content. As the authors of \cite{cim02d} pointed out, in toroidal compactifications this model must be seen as a {\it local} construction, where extra R-R sources 
such as hidden sector branes and/or background fluxes should be added. 
Several attempts have been made to embed this local construction into
a global model. First,  it was shown in \cite{dhs04} that 
this local model can be embedded into an abstract conformal field theory construction known as
Gepner orientifold (see subsection \ref{ssgb}). These Gepner constructions
are  located at special points in the Calabi-Yau moduli space where the geometric intuition is lost,
\footnote{For instance, it is not straightforward how to introduce background fluxes.} 
and so it is therefore desirable to find an embedding into a geometrical construction.
However, it proves difficult to do so for a  toroidal (orbifold) background without 
introducing anti-branes because the cancellation of R-R tadpoles requires
{\it some} R-R charges of a D-brane to have the same sign as that of an O6-plane.
Peculiar as it might seem, it was pointed out in an earlier  work \cite{csu01,csu01a} that
D-branes with this property do exist. Armed with this observation, 
two independent attempts \cite{clll04} and \cite{ms04,ms04a}
were made to embed the local model of \cite{cim02d} into a $\mbb{Z}_2 \times \mbb{Z}_2$
 orientifold, and indeed a consistent global realization of \cite{cim02d}
was found in \cite{ms04,ms04a}.
Unfortunately, in the original version of \cite{clll04},
only the homological R-R charges are canceled but the K-theory constraints
\cite{ms04a} are not satisfied,  resulting in 
the massless spectrum with  discrete global anomalies \cite{EW82}.
In the revised version of  \cite{clll04},  employing  the K-theory 
constraints derived in \cite{ms04a}, a consistent model was obtained 
with minor modifications. The Higgs sector of 
the model  is  no longer minimal, unlike the
construction in \cite{ms04,ms04a}. It should be noted that the hidden 
sector D-branes introduced in these global models \cite{ms04,ms04a,clll04} 
have non-trivial intersections with the Standard Model sector D-branes and 
so there are chiral exotics.

\subsection{Supersymmetric models on more general  backgrounds}
\label{ssgb} In the recent years other supersymmetric intersecting D-brane
models have been constructed with the aim to find realizations of the MSSM.
Essentially, two different classes of string backgrounds were considered.
First, using the methods reviewed in section \ref{soib}, 
more complicated orbifold backgrounds like a $\mbb{Z}_4$, $\mbb{Z}_4\times
\mbb{Z}_2$  or $\mbb{Z}_6$ orbifold have been studied. 
Second, to move beyond toroidal orbifolds
and to consider intersecting branes on more general Calabi-Yau spaces, methods
to treat Gepner model orientifolds were developed. 
Let us briefly review these activities in the
following two sections.

\subsubsection{Other toroidal orbifolds}

\label{sssto} One way to generalize the $\mbb{Z}_2\times \mbb{Z}_2$
orientifolds studied above is to include additional shift symmetries in the
$\mbb{Z}_2$ actions \cite{GP02,GP03,LG03}. These have the effect of eliminating
some of the orientifold planes present in the original models, which would
make it much harder to find interesting supersymmetric models. On the other
hand, it also gives rise to twisted sector three-cycles, which allows for more
general fractional D6-branes. Some of these models were constructed in
\cite{GP02,GP03,LG03}.

Employing the topological methods introduced in  section \ref{soib}
\cite{bbkl02}, chiral supersymmetric intersecting D-brane models have been
studied so far on the $\mbb{Z}_4$ \cite{bgo02}, $\mbb{Z}_4\times \mbb{Z}_2$
\cite{GH03,GH03a,GH04} and $\mbb{Z}_6$ \cite{ho04} toroidal orbifolds. In the
first two cases, it turned out that
 semi-realistic MSSM-like models could only be achieved after certain D-brane recombination
processes were taken into account (see the original papers for more details).
For the $\mbb{Z}_6$ model \cite{ho04}  the authors were performing an
exhaustive search for MSSM-like models and found a class of interesting
D-brane configurations, which gave rise to the MSSM spectrum without the
complication of brane recombinations.

Moreover, there are  both four and six-dimensional toroidal backgrounds, where
so far only the non-chiral solutions to the tadpole cancellation condition,
with D6-branes placed parallel to the orientifold planes, have been considered
\cite{bgk99,GP99,ab99,bgk99a,bgk00,fhs00,bgkl00,bcs04}.

\subsubsection{Gepner Model orientifolds}
\label{sssgmo} One of the unattractive phenomenological features of all the
toroidal orbifold models discussed above is that they give rise to too many
adjoint scalars. Geometrically this means that the three-cycles  $\pi_a$  one
is considering have too many deformations, which are counted by $b^1(\pi_a)$.
Not only for this reason, it is desirable to have many more backgrounds
available. However, for more general algebraic Calabi-Yau spaces not very much
is known about sLag three-cycles, which prevents a direct geometric approach
to the problem as pursued for instance in \cite{bbkl02,AU02,AU03a,bbkl02a}.

One way out is to use Gepner models, which are exactly solvable conformal
field theories known to describe certain symmetric points in the moduli space
of distinguished Calabi-Yau manifolds. Since the description of D-branes and
orientifold planes in this context is a subject of its own, here we would like
to only mention that after some first attempts \cite{abpss96,bw98}
during the last few years methods have been  developed to
treat Gepner model orientifolds very efficiently
 \cite{aaln03,RB03a,bhhw04,bw04,dhs04,aaj04,bw04a} allowing
a systematic computer search for MSSM like models. Specifically, the
impressive results of \cite{dhs04a} provide large classes of three-family
Standard-like Models with no chiral exotics. It remains to be seen
whether  some of these models also satisfy the more refined
Standard Model constraints.
Note however that these
exact conformal field theory models are located at very special points in  the
K\"ahler and complex structure moduli space where the geometric intuition is
lost.  Since one expects that all radii are of string scale
size, couplings,
such as Yukawa couplings, are not expected to  possess hierarchies
associated with the size
of the internal spaces, such as in the case of the toroidal orbifolds with
D-branes. In addition, the
introduction of supergravity fluxes is not straightforward to perform
(see section \ref{ssbf}).

\section{LOW ENERGY EFFECTIVE ACTIONS}
\label{sleea}

The models presented in the previous section  \ref{ssrm} provide a starting
point for  the study of couplings in the effective low energy theory whose
massless spectrum was determined by techniques presented in section
\ref{soib}. For the
 orientifold models with intersecting  D6-branes compactified on orbifolds,
the calculation of such couplings can be done by employing the conformal field
theory techniques on orbifolds \cite{dfms87}.  The tree level calculations
can in principle be performed both for the supersymmetric and the
non-supersymmetric constructions; for the one-loop calculations supersymmetry
is a necessary ingredient to  obtain an unambiguous  finite answer. The
summary of the explicit results will therefore focus primarily on the
supersymmetric constructions (Alternatively, part of the low energy effective
action can also be determined by a dimensional reduction of the
ten-dimensional supergravity theory \cite{gl04,jl04,gl04a}).

The calculation of  couplings for chiral superfields  at the  intersection of
D6-branes  and,  in particular, the  Yukawa couplings for  such states are
clearly of phenomenological interest. As discussed in section \ref{ssc}
the states at such intersections
correspond to the open string excitations stretched between the two
intersecting D6-branes. As a consequence
 the bosonic  string oscillator modes
are  like given in  eq.(\ref{eom}).
Therefore, physical string excitations at the two D6-brane
intersections are associated with the twisted open-string sectors, which are
analogous to  the  closed string twisted sectors  on orbifolds \cite{dfms87}.

The string amplitudes for  these  excitations can in turn be calculated
employing conformal field theory techniques \cite{dfms87}. The correlation
functions of the fermionic string excitations   can be obtained in a
straightforward manner by employing a world-sheet bosonisation procedure. On
the other hand the correlation functions for  the bosonic excitations  involve
the calculation of the correlation functions for the  so-called bosonic twist
fields, i.e., $\sigma_{\epsilon_i}(x)$, evaluated  at the world-sheet
location $x$ on the disc. The bosonic twist field ensures  that the bosonic
open string  fields $X^i(z)$  (in  the i-th toroidal direction)  have the
correct twisted boundary conditions. Here $z$ is the world-sheet coordinate.
These boundary conditions are  encoded in the following operator product
expansion \cite{dfms87,cp03}:
 \begin{eqnarray}
 \label{OPEs}\partial
X^i(z)\sigma_{\epsilon_i}(x)&\sim&(z-x)^{{\epsilon_i}-1}
\tau_{\epsilon_i}(x)+\ldots \nonumber
\\ \nonumber
\partial
\tilde{X}^i(z)\sigma_{\epsilon_i}(x)&\sim&(z-x)^{-{\epsilon_i}}
\tau'_{\epsilon_i}(x)+\ldots
\\ \nonumber
\bar{\partial}X^i(\bar{z})\sigma_{\epsilon_i}(x)&\sim&-(\bar{z}-x)^{-{\epsilon_i}}\tau'_{\epsilon_i}(x)+\ldots
\\
\bar{\partial}\tilde{X}^i(\bar{z})\sigma_{\epsilon_i}(x)&\sim&
-(\bar{z}-x)^{{\epsilon_i}-1}\tau_{\epsilon_i}(x)+\ldots
\end{eqnarray}
 and similarly for $\sigma_{-{\epsilon_i}}(x)$. Here
$\tau_{\epsilon_i}$ and $\tau'_{\epsilon_i}$ correspond to the
excited bosonic twist fields.  Employing the so-called stress-energy method
\cite{dfms87}, which allows one to determine the correlation functions of
bosonic  twist fields  by employing the properties of the operator product
expansion of the conformal field theory stress-energy tensor with the twist
fields, along with the above operator  product expansions (\ref{OPEs}),
enables one \cite{cp03} to  determine the bosonic twisted sector string
amplitudes.
The application of these calculations to the four-point couplings and Yukawa
couplings  will be discussed in subsection \ref{sssyc}.

Another set of  couplings involves the calculation of the  K\"ahler potential
for the states at the intersection, in particular the explicit dependence of
the  leading term which is bi-linear in powers of chiral superfields at the
intersection. The corresponding string amplitudes involve the correlation
functions containing both   states at the intersection (open string states)
and toroidal moduli fields (closed string states) \cite{lmrs04}. Such
couplings will  be discussed in subsection \ref{ssskp}.

Another important topic is  the   calculation of the gauge couplings. In
particular,  determination of the  holomorphic gauge kinetic function  in terms of the
dilaton and the toroidal moduli  both at the tree  and one-loop level is an
important task and will be discussed in section \ref{ssgc}.

 \subsection{Correlation functions for  states at D-brane intersections}
\label{sstf}

In this section we summarize the results for the tree level calculations for
the chiral matter  appearing at the D6-brane  intersections (see section
\ref{ssc}).
 In  the supersymmetric
constructions the states at intersections correspond to the  full  massless
chiral supermultiplet.  The couplings of most interest are the tri-linear
superpotential couplings, such as the coupling of quarks and leptons to the
Higgs fields. On the other hand the four-point couplings are also of interest,
since they indicate the appearance of  higher order terms in the effective
Lagrangian; for example, certain four-fermion couplings could contribute  to
the  flavor changing neutral currents in the Standard-like model constructions
(see \cite{ao03,ams03}) and in the Grand Unified Models  triggering proton  decay (see
\cite{kw03}).

\subsubsection{The four-point and three-point  functions - Yukawa couplings} 
\label{sssyc}

The explicit calculations of the three-level four-point and three-point
correlation functions for the states appearing at the  D-brane intersections
 were done in \cite{cim03,cp03,ao03,kw03}. Generalizations to n-point functions were
addressed in  \cite{ao03a}.

As discussed in the introduction of this  section the non-trivial part in the
calculation  involves the evaluation of the correlation functions of four
(three) bosonic twist fields, which  signify the  fact that the  states at the
intersection arise from the sector with twisted boundary conditions on the
bosonic and fermionic string fields.  The conformal  field  theory techniques
employed  are related to the study of bosonic twist fields of the closed
string theory on orbifolds \cite{dfms87}.  For technical details of the
specific calculation of the four- and three-point functions, employing 
conformal field theory techniques,  we refer the
reader to refs. \cite{cp03,lmrs04}, and for a detailed calculation of the 
classical part of Yukawa couplings to ref. \cite{cim03}.

The calculations have been done in the case of  intersecting D6-branes
wrapping factorizable three-cycles of a six-torus $T^{6}=T^2\times T^2\times
T^2$. Thus, in each $T^2$ the D6-branes wrap one-cycles, and the problem
reduces to a calculation of  correlation functions of bosonic twist fields
associated with the twisted sectors at intersections of D6-branes  wrapping
the one-cycles of a $T^2$. 
The final answer is therefore a product of
contributions from correlation functions on each of the 
three $T^2$ \cite{cim03}.

In particular,  the following  four-point correlation functions of bosonic
twist fields are of interest:
\begin{equation} \langle \sigma_\nu (x_1)\sigma_{-\nu} (x_2)\sigma_\nu
(x_3)\sigma_{-\nu} (x_4)\rangle \label{nunu}
\end{equation}
and
\begin{equation}
\langle\sigma_\nu (x_1)\sigma_{-\nu} (x_2)\sigma_{-\lambda}
(x_3)\sigma_{\lambda} (x_4)\rangle. \label{nulambda}
\end{equation}

\begin{figure}
\begin{center}
\scalebox{0.5}{\rotatebox{-90}{\includegraphics{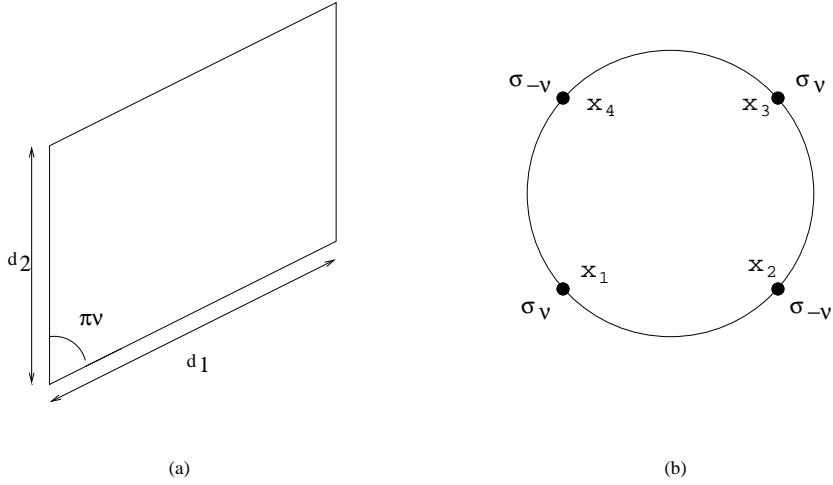}}}
\end{center}
\caption{\small  Target space: the  intersection of two parallel branes
separated by  respective distances $d_1$ and $d_2$ and intersecting at angles
$\pi \nu$ (Figure a).   World-sheet:
 a disk diagram of the
four twist fields located at $x_{1,2,3,4}$ (Figure b). The calculation
involves a map from the world-sheet to target space. } \label{ft}
\end{figure}

The first one corresponds to the bosonic twist field correlation function of
states appearing at the intersection of two pairwise parallel branes with
intersection angle $\pi\, \nu$ (see Figure \ref{ft}). This correlation
function is a key ingredient in the calculation of the four-fermion couplings,
that contributes  to the flavor changing neutral currents in the Standard-like
models \cite{ams03} and to the proton  decay  amplitudes in the Grand Unified
Models \cite{kw03}.

\begin{figure}
\begin{center}
\scalebox{0.5}{\rotatebox{-90}{\includegraphics{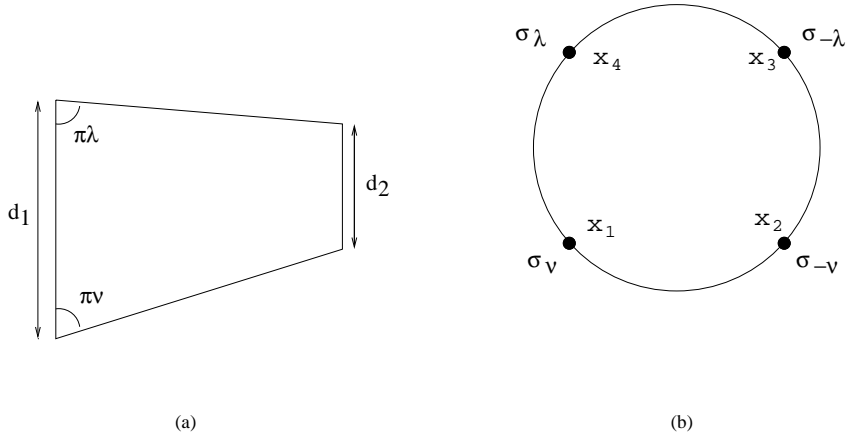}}}
\end{center}
\caption[]{\small  Target space: the  intersection of two branes intersecting
respectively with the two parallel branes at angles $\pi \nu$ and $\pi
\lambda$, respectively (Figure a). World-sheet: a disk diagram of the four
twist fields located at $x_{1,2,3,4}$ (Figure b). The calculation involves a
map from the world-sheet to target space, allowing for a  factorization to a
three-point function. } \label{fta}
\end{figure}

The second amplitude (\ref{nulambda}) corresponds to the bosonic twist field
correlation  function of states appearing at the intersection of two  branes
intersecting at respective angles $\pi\nu$ and $\pi \lambda$ with the third
set of parallel  branes (See Figure \ref{fta}). This correlation function is
specifically suited for  taking the limit of the world-sheet coordinate
$x_2\to x_3$  which factorizes to a three point function associated with the
intersection of three branes. This latter result is particularly interesting
since it provides a key element in the calculation of the Yukawa coupling.

By employing the stress-energy conformal field theory techniques and the
properties of the operator product expansions of the bosonic twist fields
(\ref{OPEs}) one can determine \cite{cp03}  both the classical part and the
quantum part of  such  amplitudes and thus the exact tree level answer for the
corresponding couplings. The calculation of the quantum part depends only on
the intersection angles and is  thus insensitive to the  scales of the
internal space and relative position of the branes.
 On the other hand the classical part carries information on the actual
separation among the branes  and the overall volume of $T^2$ as well.  An
important part in the calculation is the determination of an overall
normalization of the four-point  amplitude, which  can be
    done by factorizing the amplitude in the limits $x\to 0$ or
$x\to 1$, where it reduces to  a product of the two
three-point
    amplitudes. Namely, in these limits, the four-point amplitude
     contains a dominant  contribution from the exchanges of the
    intermediate open string winding states around the compact
    directions. The dominant contribution  can be interpreted as the
  $s$ (or $t$)-channel exchange of the massless
gauge bosons living on the world-volume of the D6-branes. Thus,
 in this limit the amplitude is completely determined in terms of the
gauge-coupling which in turn  determines  the normalization of the full
four-point amplitude. For details see \cite{cp03,kw03}.

We shall skip the technical details and in  the following  only quote the
result for the exact (string) tree-level expression for the Yukawa coupling
for two massless fermionic states and one massless 
bosonic  state, appearing at the
intersections of three D6-branes \cite{cp03}:
\begin{equation}
Y=\sqrt{2}g_0 2\pi\prod_{j=1}^3\left[\frac{16\pi^2 \Gamma(1-\nu_j)
\Gamma(1-\lambda_j)\Gamma(\nu_j+\lambda_j)}{
\Gamma(\nu_j)\Gamma(\lambda_j)\Gamma(1-\nu_j-\lambda_j)}\right]^\frac{1}{4}
\sum_{m\in\{{\rm ws-inst.}\}}\exp\left( -\frac{A_j(m)}{2\pi\alpha'} 
\right) \, , \label{yuk}
\end{equation}
where $A_j(m)$ is the area of the m-th triangle (world-sheet instanton)
formed by the three
    intersecting branes on the j-th two-torus and  $g_0=e^{\Phi/2}$,
    with $\Phi$ corresponding to the Type IIA dilaton.
In order to derive (\ref{yuk}) it was assumed that all three fields have
canonically normalized kinetic energies. For a discussion of  phenomenological
implications the above results,  see subsection \ref{Yukawaunif}.

In  earlier works  \cite{cls02a,cim02d} the leading classical contribution to
such Yukawa couplings was calculated. Also previously, a comprehensive analysis and computation of the 
full classical contribution (which contains all the open string 
moduli dependence)
was performed in \cite{cim03}.
Intriguingly, in
\cite{cim04}  the three-point function was calculated  in the mirror-dual Type
IIB theory.  Here a purely classical, leading order in $\alpha'$, computation
gave  already the full world-sheet instanton corrected Type IIA superpotential
contribution to the Yukawa couplings (see eq.(\ref{yuk})). This can be
considered as a nice confirmation of mirror  symmetry. As expected the
K\"ahler potential contributions to the Yukawa couplings only agreed to 
leading order in $\alpha'$.

The prefactor in  (\ref{yuk}) corresponds to the quantum part of the bosonic
twist correlator; it  has a suggestive factorizable form associated 
with the
angles of states appearing at each intersection, resulting in the 
following  complex 
structure dependence of the K\"ahler potential for chiral superfields 
at D6-brane intersections:
\begin{equation}
K= \frac{1}{4\pi}\, \sum_{\nu} \prod_{j=1}^3 \sqrt{
\frac{\Gamma(\nu_j)}
{\Gamma(1-\nu_j)}} \, \Phi_\nu\, \Phi^*_\nu, \nonumber\, .
\end{equation}
(For simplicity in the above expression the Planck scale
was set to $M_{P}=1$.)
  On the  other hand, the
classical part of the Yukawa coupling
is proportional to the  area of the
intersection triangles and thus depends on the toroidal K\"ahler structure 
moduli.  
It includes a contribution from  
the part of the K\"ahler  potential  that depends on toroidal K\"ahler 
moduli   \cite{cim04}. 
Thus the full K\"ahler potential takes the form displayed
in the next subsection (see eq.~(\ref{Kahler})).
 After the inclusion of 
 toroidal  two-form field potentials and the asscoiated Wilson lines 
\cite{cim03}, 
the remaining part of the Yukawa coupling
describes  the superpotential tri-linear coupling as a  holomorphic 
function of toroidal K\"ahler 
moduli; this coupling typically takes a form of 
modular theta functions  (for further details, see 
\cite{cim03,cim04}). 
This splitting of the three point function (\ref{yuk}) into the leading
K\"ahler potential contribution  and  the superpotential contribution
has been confirmed as a part of the  calculations described in the
following subsection.

Higher tree level n-point correlation functions for chiral superfields
at D-brane intersections have been studied in \cite{ao03,ao03a}. The one-loop
calculation of the three point functions for such states  was done in
\cite{as04} and it leads to new results for the one-loop corrections to the
K\"ahler potential for  the corresponding chiral superfields at D-brane
intersections.

\subsubsection{The closed-open string amplitudes - K\"ahler potential}
\label{ssskp}
A direct  calculation of  the tree level leading order K\"ahler potential for
chiral superfields  at D-brane intersections and their dependence  on the
closed string sector moduli involves the determination of the  string
amplitudes for the   two  open string sector  vertices and an arbitrary number
of closed sector moduli vertices. For the toroidal (orbifold) backgrounds
these calculations have been  carried out  explicitly for any number of
toroidal complex and K\"ahler structure  moduli in \cite{lmrs04}.
First   explicit results for the four-point string amplitudes were derived and
then by further employing the symmetry structure of higher n-point functions,
sets of  differential equations were obtained for the n-point functions
with an arbitrary number of  vertices for the  toroidal moduli.
These  could  be
explicitly solved, thus resulting in the explicit string amplitudes with any
number of toroidal closed sector moduli. As a consequence, the
leading  order tree level K\"ahler potential for chiral superfields at D-brane
intersections, and its explicit dependence on {\it both} the toroidal K\"ahler
and complex structure moduli could be derived.
Specifically, the K\"ahler  potential  for the
open string sector chiral superfields $\Phi_{\nu_{ab}}$, appearing at the
intersection of the stack $a$ and stack $b$ of D6-branes, takes the form
\cite{lmrs04}: \begin{equation}
K = \frac{1}{4\pi}\, \left[\prod_{i=1}^3 (T_i +T^*_i)^{-\nu^i_{ab}}
         \, \sqrt{ {\Gamma( \nu_{ab}^i)\over
              \Gamma(1- \nu_{ab}^i)}}\right] \Phi_{\nu_{ab}}\Phi^*_{\nu_{ab}}\, ,
\label{Kahler} \end{equation} where again $\pi\, \nu_{ab}^i$ denotes the angle
of the $a$- and $b$- D6-brane intersection in the $i$-th two-torus  and $T_i$
is the K\"ahler modulus of the $i$-th two-torus. (For simplicity again the Planck
scale is set to $M_{pl}=1$.) Note that the dependence of the above K\"ahler potential
on the angles and thus implicitly on the toroidal complex structure moduli
is the same as the one obtained from the Yukawa coupling calculation
(\ref{yuk}). In addition (\ref{Kahler}) also contains the information on 
the
toroidal K\"ahler moduli.

\subsection{Gauge couplings}
 \label{ssgc}
 The last function that specifies the effective  four-dimensional ${\cal N}=1$ supersymmetric theory is the
 gauge kinetic  function. The tree level gauge kinetic function for each
 stack of D6-branes  can be determined in a straightforward way
  by reducing the D6-brane world-volume
 kinetic energy action along the three-cycle
 wrapped by the stack of D6-branes in the internal space.
 For a supersymmetric  three-cycle,
 $\pi_a$, the tree level  gauge kinetic function is a holomorphic function
 of  complex structure moduli  fields  and it is of the form \cite{cim02,bbkl02,bls03}
\begin{equation}
   f_a={M_s^3\over (2\pi)^4}\left[ e^{-\varphi} \int_{\pi_a}
     \Re (\Omega_3) +2i \int_{\pi_a} C_3 \right]\, ,
\end{equation}
where $C_3$ denotes the R-R three-form.
For supersymmetric three-cycles on toroidal orbifolds this  holomorphic gauge
kinetic function takes an explicit form in terms of
the toroidal complex structure moduli
$U^{i}$ and the dilaton field $S$
\bea
     S&=& {M_s\over 2\pi} e^{-\varphi}\, \prod_i R_1^i + {i\over 4\pi} C^0\, , \\
     U^i&=& {M_s\over 2\pi} e^{-\varphi}\,  R_1^i\, R_2^j\, R_2^k + {i\over
       4\pi} C^i \, , \nonumber
\eea
with $i\ne j\ne k\ne i$.
For example in
the case of the $\mbb{Z}_2\times \mbb{Z}_2$ orientifold,  the  gauge
coupling function  takes the form (see, e.g., \cite{cim02, clw03}):
 \begin{equation}
   f_a(U^i,S)=\textstyle{1\over 4}\left[n^1_a\,n^2_a\, n^3_a\, S\,  -\,
n^1_a\,{\widetilde m}^2_a\, {\widetilde m}^3_a\, U^1\,
    -\, {\widetilde m}^1_a\,{ n}^2_a\, {\widetilde m}^3_a\, U^2\,  -\,
     {\widetilde m}^1_a\,{\widetilde m}^2_a\, {n}^3_a\, U^3\,
     \right]\, ,\label{tgc}
 \end{equation}
 where  as usual $n^i_a$ and ${\widetilde m}^i_a$
 are the wrapping numbers of  the three-cycle
 $\pi_a$ and  the pre-factor $\textstyle{1\over 4}$ is the dimension of
the orbifold/orientifold
 group.

 We  emphasize that since the gauge coupling  for  each gauge group factor
 depends on the  volume of the corresponding three-cycle $\pi_a$,
  in general the intersecting D-brane constructions do not have  gauge coupling
 unification in the sense of Grand Unified Models.
However, for each gauge group factor the
 tree level gauge coupling is calculable in terms of the toroidal complex
 structure  moduli, the dilaton  and the wrapping numbers of the three-cycle $\pi_a$.
The results for the  tree level gauge kinetic functions were employed
 as the starting point to  address the renormalization group running of gauge couplings
  from the string
 scale  to the electroweak regime for the semi-realistic constructions
 \cite{cls02,bls03,RB03} as well as in the study of the moduli stabilization
and  supersymmetry breaking due to  gaugino condensation in the hidden sector
 of supersymmetric semi-realistic constructions \cite{clw03} (for the
 respective
 phenomenological implications see subsections  \ref{group} and \ref{sssgd}).

The tree level couplings receive corrections at the one-loop level due to the
so-called  threshold corrections of the heavy string modes.  These explicit
calculations are involved, since the complete massive string spectrum is
needed. For  the perturbative heterotic string theory on orbifolds these
threshold corrections were first calculated in \cite{dkl90}. The calculation
of the gauge coupling threshold corrections for the  intersecting D6-branes on
a toroidal orbifold backgrounds amounts to  similar complexity and the
explicit results have been computed in \cite{ls03}. While the one-loop
corrections for open string sectors preserving ${\cal N}=4$ supersymmetry
vanish, the ${\cal N}=2$  sectors  depend on both the toroidal  complex and
K\"ahler moduli. For explicit expressions please consult \cite{ls03}.
These corrections bear similarities with  the  heterotic
orbifold corrections \cite{dkl90}. For details and specific explicit
calculation for the $\mbb{Z}_2\times \mbb{Z}_2$ orientifold, see again
\cite{ls03}. In
addition, there are ${\cal N}=1$ sector corrections; they  can be cast in a
compact expression which for the $SU(N_a)$ gauge couplings takes the form:
\begin{equation}
  \Delta_{ab}=-b_{ab} \ln{\Gamma (1-\nu_{ba}^1)
        \Gamma(1-\nu_{ba}^2)
          \Gamma(1+\nu_{ba}^1+\nu_{ba}^2)\over
         \Gamma(1+\nu_{ba}^1)
        \Gamma(1+\nu_{ba}^2)
          \Gamma(1-\nu_{ba}^1-\nu_{ba}^2)}\, ,
\end{equation} where $b_{ab}=N_b I_{ab} {\rm Tr}(Q^2_a)$ and
$\pi\, \nu^i_{ba}$ denote    the intersection angles  of a and b D6-branes in
the i-th two-torus. These angles can be expressed in terms of the wrapping
numbers  of the $\pi_a$ and $\pi_b$ three-cycles and the toroidal complex
structure moduli $U^i$. (For the explicit formula, see, e.g., \cite{cp03}.)
The total correction from the ${\cal N}=1$ sector is obtained as a summation
over all $b$'s, i.e., stacks of all the other D6-branes wrapping the
three-cycles $\pi_b$.

These threshold corrections could play an important role in the  study of the
renormalization group running of gauge couplings; in particular, they can
modify the effective string scale.  In addition, since the threshold
corrections depend on both the K\"ahler and complex structure moduli they
could play an important role in the strong infrared  ``hidden sector''
dynamics, and
the possibility of stabilizing all  toroidal moduli.

\section{FLUX VACUA WITH MAGNETIZED \- D-BRANES}
\label{ssbf}

As we have seen in the previous sections, intersecting D-brane worlds provide
a simple geometric framework within which many semi-realistic particle physics
models can be constructed. However, just like other supersymmetric string
constructions, these intersecting D-brane models suffer from the usual moduli
problem. Typically, these models contain a lot of moduli (from both closed and
open string sectors) which remain massless before supersymmetry is broken and
hence, if not stabilized, would lead to serious phenomenological problems as
well as loss of predictivity. Deeply related to the moduli problem is the
question of how supersymmetry is broken. In studying the phenomenological
consequences of string theory, one traditionally starts with an ${\cal N}=1$
supersymmetric string vacuum whose low energy spectrum contains the Standard
Model. The hope is that the same mechanism that breaks supersymmetry and gives
masses to the superpartners of the Standard Model could also lift all the
moduli. 
Strong D-brane gauge dynamics, for example, can result in gaugino and matter
condensations, generating a non-perturbative Veneziano-Yankielowicz-type
potential  \cite{vy82,vty83} that can in principle stabilize certain
closed string moduli. However, such  non-perturbative dynamics
can  be analyzed only at the level of an
effective super Yang-Mills theory with the leading instanton contribution.
 Moreover, without fine-tuning, the vacua stabilized by
non-perturbative effects typically have a large non-vanishing cosmological
constant, thus rendering these models unrealistic for further phenomenological
studies (see section \ref{susybreaking}). Therefore, in practice, one often
treats the supersymmetry breaking sector as a black box and simply
parameterizes our ignorance of supersymmetry breaking with the VEVs of the
auxiliary fields of some moduli without specifying how they acquire a VEV
(see, e.g., \cite{bim97}).

Recently, there have been some interesting attempts to understand
simultaneously these
two central problems in string phenomenology -- moduli stabilization and supersymmetry breaking
 -- by considering compactification with background
flux \cite{gvw99,drs99,tv99,gss00,cklt00,gkp01}.
The idea, which is most conveniently expressed in the framework of
Type IIB string theory, is that
 a superpotential can be generated by the
NS-NS and R-R three-form flux background \cite{gvw99,tv99}. The superpotential
thus generated depends on the dilaton and the complex structure moduli, and so
these moduli are generically lifted. Interestingly, depending on how the gauge
and chiral sectors are embedded, supersymmetry can be {\it softly} broken by
the flux. More importantly, the resulting soft SUSY breaking terms can be
calculated in a systematic way {\it perturbatively}
\cite{ciu03,ggjl03,lrs04,lrs04a,ciu04}. We shall see towards the end of this
section how such soft terms are generated from the local D-brane physics point
of view. From the effective four-dimensional supergravity point of view, the
effect of the background flux is to introduce a {\it microscopic} source for
the auxiliary fields of the moduli which as a result breaks supersymmetry.

Thus, flux compactification provides a rather attractive framework for string
phenomenology. However, in order to explore {\it quantitatively} its
phenomenological features, it is important to construct some concrete examples
in which realistic features of the Standard Model, such as chirality, can be
incorporated. The general techniques of constructing chiral flux vacua have
been developed in \cite{blt03,cu03,cu03a}, although no chiral models which are
free of tadpole instability have been found. More recently, it was realized in
\cite{ms04,ms04a} that a crucial ingredient in constructing stable chiral flux
vacua is to introduce additional pairs of $D9-\overline{D9}$-branes which
nonetheless are BPS because of the magnetic flux on their world-volumes.
Chiral flux vacua (both supersymmetric and non-supersymmetric) that are free
of tadpoles have been constructed in \cite{ms04,ms04a}. Furthermore, the low
energy spectrum of these models is remarkably close to the MSSM, and hence
they provide a proof of concept that realistic particle physics features can
be embedded in flux compactification \footnote{{\it Local} models of flux
compactification with realistic particle physics features have been considered
in \cite{cgqu03}.}. Interestingly, in cases where supersymmetry is broken
softly by the background flux, the vacuum remains Minkowski after
supersymmetry breaking (at least to lowest order) since the NS-NS tadpoles are
absent. Subsequently, there have been further interesting attempts in
constructing realistic models within this framework and more examples of 
three- and four-family chiral flux vacua, including supersymmetic ones, 
have been found in \cite{cl04,cll05}.

Although this review has so far been focused on Type IIA string theory with
intersecting D6-branes, the intersecting D-brane models discussed here are
related (in the absence of flux) by a simple duality to Type IIB orientifolds
with magnetized D9-branes (see, e.g., \cite{cu03} for the details of such
map). Hence, there is an alternative, albeit less geometrical, description of
the same models in Type IIB string theory. In fact, the techniques of building
intersecting D-brane models that we have discussed can be readily adapted to
construct chiral flux vacua, which we will review below.

\subsection{Three-form fluxes in Type IIB string theory}

Various aspects of flux compactifications have been
discussed in the literature.
Instead of providing a comprehensive overview
of flux compactifications (which is not the main purpose of this review),
we will only sketch here some of
the basic results \cite{drs99,gkp01,kklt03,kstt02}
relevant to string model building.
Consider Type IIB string theory in the presence of a non-trivial three form
background flux $G_3 = F_3- \tau H_3$.
Here $F_3$ denotes the R-R
and  $H_3$ the NS-NS three form flux, and $\tau$
is
the complex dilaton-axion field.
The background fluxes must obey
the Bianchi identity and be properly quantized, i.e.,
they take values in $H^3({\cal M},\mbb{Z})$.
In toroidal (orbifold) backgrounds, such
quantization conditions are particularly simple:
 \begin{equation}
    {1\over (2\pi)^2 \alpha' }  \int_{\Sigma} H_3\in N_{min}\times \mbb{Z}, \quad\quad
    {1\over (2\pi)^2 \alpha' }  \int_{\Sigma} F_3\in N_{min}\times  \mbb{Z},
\end{equation}
 where $\Sigma$ is a three-cycle in the {\it covering space},
$N_{min}$ is a positive  integer which reflects the fact that in an
orientifold (orbifold), there can exist three-cycles which are smaller that in
the covering space.
 For example for the   $T^6/\mbb{Z}_2\times \mbb{Z}_2$ orientifold, taking into account
also the orientifold projection,
  $N_{min}=8$, and $N_{flux}$, defined in (\ref{tadpoleb}), is a multiple of 64.

Turning on   $F_3$ and $H_3$ fluxes  has two important effects.
First, the Chern-Simons terms in  the Type IIB effective supergravity action
\be
     S_{CS}={1\over 2\kappa_{10}^2} \int d^{10} x {C_4\wedge G_3\wedge \overline G_3\over
      4 i\, {\rm Im}\tau}\, ,
\ee when integrated over the six-dimensional manifold $\cal M$ induces a
tadpole for the R-R four-form gauge potential $C_4$. In particular, the D3
charge contribution to the tadpoles is of the form
\begin{equation}
N_{flux}={1\over (4\pi^2\alpha')^2 } \int_{\cal M} H_3\wedge F_3\,
.\label{tadpoleb}
\end{equation}
The second effect is that the kinetic term for $G_3$ (suppressing the
warp factor)
\be
     V={1\over 4\kappa_{10}^2 {\rm Im}\tau  } \int_{\cal M} d^{6} y \,
     G_3\wedge \star_6 \overline G_3\, ,
\ee
induces a scalar potential, which can be written as
\be
\label{potent}
     V={1\over 2\kappa_{10}^2 {\rm Im}\tau} \int_{\cal M} d^{6} y \,  G^-_3\wedge \star_6 \overline G^-_3 -
      {i\over 4\kappa_{10}^2 {\rm Im}\tau}  \int_{\cal M} d^{6} y \,  G_3\wedge  \overline
      G_3\, ,
\ee
where again we suppress the warp factor.
Here, $G^{\pm}_3$ is the imaginary self-dual/anti self-dual (ISD/IASD)
part of $G_3$, i.e., it satisfies
$\star_6  G^{\pm}_3=\pm i G^{\pm}_3$.
The second term in (\ref{potent})
is a topological term equal in magnitude to $N_{flux}$
and gives rise to the NS-NS tadpole
of the flux. Contrarily, the first term in
(\ref{potent}) is a positive semi-definite  F-term potential \cite{gvw99,tv99}, which precisely vanishes
if the flux is imaginary self dual, i.e., $G^-_3=0$.

It has been shown in \cite{tv99} that this F-term potential $V_F$ can be derived from the
Gukov-Vafa-Witten superpotential \cite{gvw99}
\begin{equation}
\label{gvw}
W=\int_{\cal M} \Omega_3\wedge G_3\, ,
\end{equation}
which  apparently depends only on  the complex structure moduli and the dilaton
and  vanishes if these moduli are chosen such that   $G_3$
is imaginary self-dual. Self-duality implies
that the three-form flux has a $(2,1)$ and a $(0,3)$ component
with respect to the complex structure of the underlying Calabi-Yau, i.e.,
$G_3=G_3^{(2,1)}+G_3^{(0,3)}$.
For supersymmetric minima, one gets
the additional conditions that  (i) $G_3^{(0,3)}=0$,
and (ii) primitivity of $G_3$, {\it i.e.}, $G_3\wedge J=0$; the latter
condition is automatically
satisfied on a Calabi-Yau manifold.
Taking the back-reaction of the fluxes on the geometry into account, one
finds it is quite moderate in the (topological)
sense that one gets a warped Calabi-Yau
metric (although the metric can be strongly warped as in \cite{ks00,kklt03}).
To summarize, the flux induced  scalar potential allows one to
freeze the complex structure moduli and the dilaton at its minima.

Although the discussion here is in the context of Type IIB string theory,
there should be an alternative description in Type IIA string theory where
intersecting D6-branes (the subject of this review) can be introduced.
However, under duality, the three-form fluxes that we consider here become
metric fluxes on the Type IIA side and the underlying geometry becomes
non-K\"ahler \cite{kstt02,glmw02,ccdlmz02}. The types of three-cycles that the
D6-branes can wrap around in such non-K\"ahler geometries 
are not well
understood. Alternatively, one can study directly Type IIA orientifolds with
background fluxes (see \cite{aft03,aft03a,dkpz04} and references therein). For recent
efforts in obtaining examples of flux compactifications in massive Type IIA
supergravity with intersecting
 D6-branes, see \cite{bc03,bc04,bc04a}.

\subsection{Semi-realistic flux vacua}\label{ssfv}

In the following, we summarize the techniques for constructing
consistent Type IIB orientifolds with magnetized D9-branes
in toroidal (orbifold) backgrounds developed
in \cite{blt03,cu03}.

A  stack of $N_a$ magnetized D9-branes on toroidal orbifolds is
characterized by three pairs of
integers $(n^i_a,m^i_a)$ which satisfy
\begin{equation}
    {m^i_a\over 2\pi}\int_{T^2_i}   F_a^i = n^i_a,
\end{equation}
where the $m^i_a$ denote the wrapping numbers of the D9-brane around
the $i$-th two-torus  $T^2_i$, $n^i_a$ is the magnetic flux, and the
$F_a^i$ is the corresponding $U(1)$ magnetic field-strength on the
D9-brane. The orientifold projection  on these quantum numbers
acts as: $\Omega \overline\sigma (-1)^{F_L}:(n^i_a,m^i_a)\to (n^i_a,-m^i_a)$.
In the T-dual intersecting D6-branes
picture, these quantum numbers correspond
to the wrapping numbers $(n^i_a,m^i_a)$  of the
homology one-cycles $([a_i], [b_i])$  of the $i$-th  two-torus
$T^2_i$ that the D6-branes wrap around.
(For a detailed dictionary between these two
T-dual descriptions, see, e.g., \cite{cu03}. More general aspects
of the T-dual picture were discussed in \cite{RR01}.)

Because of the orientifold projection, the magnetized D9-branes setup as a
whole does not carry any net D5- and D9-brane charges. However, there are
additional {\it discrete} K-theory charges that needed to be taken into
account \cite{ms04a}. Other than this subtlety, the tadpole cancellation
conditions for the magnetized D9-brane sector simply amount to the
cancellation of the D3- and three types of D7-brane charges. Such conditions
can be deduced from the conditions (\ref{tadorb}) in the T-dual intersecting
D6-brane picture in section \ref{soib}. Note, however, that the background
fluxes introduce an additional contribution to the D3 tadpole
(\ref{tadpoleb}). The corresponding tadpole conditions, here specifically
written for
 the ${\mbb Z}_2\times {\mbb Z}_2$ orientifold, read \cite{blt03,cu03}:
\begin{eqnarray}
\label{tadIIB}
{\rm D3-charge} \ \ \ \ \ \sum_a N_a\, n^1_a\, n^2_a\, n^3_a &=& 8-\textstyle{N_{flux}\over 4}, \\
{\rm D7_i-charge} \ \ \ \ \ \sum_a N_a\, n^i_a\, m^j_a\, m^k_a &=&
-8\quad {\rm for}\ i\ne j\ne k\ne i \nonumber.
\end{eqnarray}
(For consistency  conditions as applied to  other orbifolds, see \cite{AF04}.)
The large positive D3-charge contribution from $N_{flux}$
makes it hard to satisfy  these tadpole conditions
without introducing anti-D3 branes which give rise to instabilities
\cite{blt03,cu03}.

Fortunately, it was realized in \cite{ms04,ms04a}
that the negative D3-brane charge needed
for the cancellation of R-R tadpoles can be accounted for by
introducing additional pairs of $D9-\overline{D9}$-branes which
are nonetheless BPS because of the magnetic flux
supported on their world-volumes.
This observation led to the first example \cite{ms04,ms04a}
 of a three-family supersymmetric Standard Model in flux compactification. In this construction, the Standard Model sector is based on the local MSSM
module introduced  in \cite{cim02d}. Therefore, as in \cite{cim02d}, there is
only a pair of Higgs doublets in the low energy spectrum and thus precisely
the minimal Higgs content of the MSSM. However, the $D9-\overline{D9}$ pairs
which carry the needed negative D3-charge have a non-vanishing ``intersection
product'' with the Standard Model building block. Hence in addition to the
gauge and matter content of the MSSM, there are also some additional chiral
exotics.

 Subsequently, additional semi-realistic flux vacua have  been
constructed in \cite{cl04,cll05}.  
In particular, 
by considering such BPS $D9-\overline{D9}$ pairs as
part of the observable sector, a broader class of
Standard Model-like vacua with three and four families of chiral matter,
and larger units of flux, including supersymmetric ones, 
were constructed in \cite{cll05}.
These models  have  typically chiral exotics and  more than one-pair of Higgs
doublets. It is fair to say that a fully realistic model of flux
compactification has yet to be found.

Let us now briefly comment on two (related) issues pertinent to these chiral
flux vacua: generation of soft supersymmetry breaking terms and stabilization
of open string moduli. First, we can understand heuristically how soft
supersymmetry breaking terms are generated by the background flux. As
discussed above, the background three-form flux carries R-R charge and
tension. More precisely, ISD (respectively IASD) flux carries the same type of
R-R charge and tension as that of a D3-brane (respectively
$\overline{D3}$-brane). Therefore, a D3-brane (respectively
$\overline{D3}$-brane) will be attracted to a region where the IASD
(respectively ISD) flux is maximum. Recall that the positions of D3-branes
correspond to world-volume scalars, and so the energy needed to move the
D3-branes away from the maximum flux region would reflect as soft masses on
the world-volume gauge theory. The analysis for the D7-brane sector is more
involved, but one can again understand how soft terms are generated by
studying the induced D3-brane charge (due to the background flux) carried by
the D7-branes.

For the same reason that soft terms are generated, the background flux
can also induce a mass to some of the open
string moduli and thus provide a way to stabilize 
them \cite{gktt04,cu04,ciu04}.
Finally, although
the K\"ahler moduli do not enter
the flux-induced superpotential,
a linear combination of
some toroidal K\"ahler moduli and open string moduli
enters the Fayet-Iliopoulos
D-term and so we expect such linear combination of closed and open string
moduli to be fixed.

Much work needs to be done before a fully realistic model of flux
compactification (i.e., a model not only with a realistic low
energy spectrum and couplings but also with all its moduli stabilized)
can be found.
However, the developments described
here have undoubtedly pointed to an interesting
direction in string phenomenology.

\section{PHENOMENOLOGICAL ISSUES}

No fully realistic intersecting D-brane model has been constructed yet. Furthermore,
many of the phenomenological features, such as the gauge and Yukawa couplings,
or the masses and other properties of the low energy
particles are model dependent or depend on details of supersymmetry breaking.
Nevertheless, it is useful to survey here some of the phenomenological
features that have emerged in various constructions, with an emphasis on the
difficulties, possibilities for new physics, and things to watch for in the future.
Many of the technical aspects or detailed consequences of specific models
were discussed in earlier sections. Here we focus on general issues.

 Let us start with two general comments. The first concerns the fundamental
string scale $M_s$. As discussed in subsection \ref{ssss}, most toroidal
(orbifold) constructions have assumed either that $M_s$ and the inverse sizes
of the extra dimensions are very large, i.e.,  within a few orders of
magnitude of
 the Planck scale, or else that $M_s$ is much lower,
e.g., in the 1- 1000 TeV range. The latter can occur  for either
supersymmetric or non-supersymmetric constructions when the overall volume of
the extra dimensions is much larger than the volumes of the three-cycles
wrapped by the D6-branes, and provides a stringy implementation of the
phenomenological brane world models with large extra dimensions \cite{add98}
(for recent reviews, see e.g., \cite{Hewett02, Lorenzana04,Csaki04}). 

Most non-supersymmetric constructions, including many toroidal ones, have
assumed a low $M_s$, of the order of the TeV scale,
to avoid large radiative corrections governed by $M_s$ which 
aggravate the Higgs-hierarchy problem, and to avoid
large contributions of  ${\cal O}(M_s^4)$  to the  cosmological constant resulting  
from  NS-NS tadpoles.
 However, as
discussed in  subsection \ref{ssss},   for the purely toroidal 
constructions with intersecting D6-branes  there is typically no direction in
the internal space that would be transverse to all the D6-branes, and thus the
size of the internal space is constrained to be of the order of the Planck
6-volume, and $M_s$ is restricted to be within a few orders of magnitude to the
Planck scale. 
The consistent implementation of the non-supersymmetric constructions
with a low $M_s$ would therefore be possible only for more general Calabi-Yau spaces
(like fractional D6-branes on toroidal orbifolds). On
the other hand for the supersymmetric constructions  the
NS-NS tadpoles cancel and  the radiative corrections
below $M_s$ are at most logarithmic. Therefore such  constructions with large
$M_s$, as dictated by toroidal (orbifold) internal spaces, are stable,
resulting in a calculable spectrum and effective Lagrangians at $M_s$.

We should also point out that  models \cite{cim02,cim02a} with locally
supersymmetric spectrum of the minimal supersymmetric Standard Model (MSSM),
which however do not cancel R-R tadpoles for toroidal orientifolds,  are also
of interest, since they may provide a prototype  D-brane configuration of the
MSSM-sector. (R-R tadpole free implementation of such construction was
realized, at the expense of Standard Model
 chiral exotics,  within ${\mbb Z}_2\times {\mbb
Z}_2$ orientifolds in \cite{ms04,clll04}; see subsection \ref{sssss}.)

Another issue to keep in mind is that  there may be new physics at the TeV
scale beyond the Standard Model or MSSM. Most of the explicit constructions
lead, e.g., to additional Higgs or exotic matter or additional $U(1)$ gauge
symmetries at the TeV scale (as do most heterotic constructions; see, e.g.,
\cite{fq2002,pl2003xa,kstv98}). It is of course possible that these are defects of
the models and that there is nothing beyond the MSSM at the TeV scale. On the
other hand, one should keep open the possibility of a rich spectrum of
``top-down'' motivated new physics, especially of the kinds that appear so
commonly in constructions.

\subsection{The spectrum}
\label{spectrum}

Most explicit intersecting D-brane constructions that contain the MSSM
spectrum also involve additional matter states. As described in section
\ref{ssrm}, it is straightforward to construct non-supersymmetric  or locally
supersymmetric intersecting D-brane models with only the Standard Model
spectrum,
 but existing fully supersymmetric
constructions are highly constrained and always contain additional matter.
This is true of most heterotic constructions as well. It is often considered
the goal of model building to come as close to the MSSM as possible, but it is
also possible that additional matter really does exist at the TeV scale. Such
states may either be chiral or non-chiral, although non-chiral states
 can typically obtain  a string scale mass after
deformations of  D-brane configurations. For  toroidal (orbifold)  models
these states involve two stacks of D-branes that  wrap the same one-cycle  in
one two-torus;  parallel splitting of  the two stacks of D-branes  in this
two-torus renders the non-chiral matter massive \cite{csu01a}.
Chiral matter occurs
in (non-Abelian) anomaly-free combinations, and
 is frequently necessary
to cancel the anomalies associated with additional gauge factors that
would not be present for the MSSM spectrum alone. There are stringent
constraints from precision electroweak physics on new matter that
is chiral with respect to $SU(2)_W \times U(1)_Y$ \cite{pdg04}
that essentially exclude the possibility of a fourth ordinary or mirror family or other
representations that are chiral under the Standard Model gauge group except possibly
for rather tuned ranges of masses or other compensations. However, states that are singlets
or vector-pairs  under the standard model group but chiral under additional gauge factors are
allowed in those cases in which  the construction allows
a mechanism for generating fermion masses in the hundreds of GeV range
(scalar masses may be due to soft supersymmetry breaking).
Additional matter may also have important implications for gauge coupling
unification, as discussed in subsection \ref{gaugeunif}.

{\em Extended Higgs Sector.} One ubiquitous possibility is an extended Higgs
sector, involving more than one pair of Higgs doublets (often many pairs) and
often standard model singlets whose VEVs could break additional gauge factors.
Additional doublets would lead to a rich Higgs spectrum detectable  at
colliders and could mediate flavor changing neutral currents. Higgs singlets
$S$ could couple to doublets with superpotential couplings $W = h S H_u H_d$,
so that a TeV-scale VEV could lead to an effective $\mu$ parameter
$\mu_{eff}=h \langle S \rangle$, elegantly solving the $\mu$ problem. This
would be an implementation of some form of the next to minimal supersymmetric
Standard Model (NMSSM) (see, for example, \cite{eghs03} and references
therein) or its $U(1)'$ extension \cite{ell02}. Such models differ
dramatically from the MSSM, e.g., by the expanded spectrum; possible large
doublet-singlet mixing with implications for Higgs masses, production, and
decays \cite{hlm04}; a different allowed range for $\tan \beta$; expanded
possibilities for electroweak baryogenesis because of a strong first order
phase transition and new sources of CP violation (see \cite{klll04,mmw04} and
references therein); and an enlarged and modified neutralino sector, extending
the possibilities for cold dark matter \cite{de97,mmw04, bkll04}. It should be
stressed that no existing construction has all of these features or all of the
couplings needed for a fully realistic model. For example, the supersymmetric
construction \cite{csu01} has no chiral singlet $S$ to generate a $\mu_{eff}$,
though its role could be played by a field in the ${\cal N}=2$ sector if it
did not acquire a large mass \cite{cls02}. (Another possibility for a $\mu$
term would be a D-brane splitting in models in which the Higgs doublets are
non-chiral, as in the locally supersymmetric model in \cite{cim02, cim02a},
although it is not clear why the splitting would be sufficiently small.)

{\em Exotic  quarks and leptons.}
Heavy exotic (i.e., with non-standard Standard Model representations)
quarks and leptons, presumably vector-like pairs  with respect to $SU(2)_W \times U(1)_Y$,
are also possible, with exotic quarks especially producing distinctive effects at a hadron collider
(see, e.g., \cite{Andre2003}).
These are familiar in $E_6$ grand unification (see, e.g., \cite{hr88}), and often emerge
in string constructions as well, e.g., associated with the remnants of a
would-be fourth family \cite{csu01}.

{\em Chiral exotics.} Still more exotic (and probably unwanted) possibilities
exist. These include the open string sector moduli, typically in the adjoint
(anti-symmetric) representation for the $U(N)$ ($Sp(2N)$) gauge symmetry, as
well as  fields in the symmetric and anti-symmetric representation for the
$U(N)$ gauge factors, associated with the intersection of D6-branes with its
orientifold image (see subsection \ref{ssms}). It is expected that the
inclusion of fluxes would  induce a  back-reaction that would give a mass to
some of the open
string moduli  \cite{cu03}. Another possibility is to introduce rigid
three-cycles like fractional branes in toroidal orbifolds.
D6-branes wrapping such three-cycles do not have massless moduli
in the adjoint representation.

Many constructions also involve intersections between the ordinary and hidden
sector branes, leading to states that are charged under the non-Abelian
factors of both, i.e., the hidden sector is really only quasi-hidden. ($U(1)$
gauge bosons also typically couple to both sectors.) These mixed states often
carry exotic electric charges (such as $1/2$). The laboratory and
astrophysical constraints on fractional charges are severe \cite{ccf96}.
Fortunately, the quasi-hidden groups are often strongly coupled, so that such
states may be confined, and may even lead to observable composite states with
more conventional quantum numbers \cite{cls02}.

{\em Grand unification exotics.} As discussed in subsection \ref{ssssgum}, it
is possible to construct Grand Unified gauge groups such as $SU(5)$ from
intersecting branes, and both fundamental and adjoint Higgs, and fundamental
and antisymmetric matter representations, appear. The supersymmetric
three-family $SU(5)$ models that have been constructed always also include
symmetric {\bf 15}-plet representations \cite{cps02},  which contain highly
exotic states such as color sextets and weak triplets. Other phenomenological
aspect of grand unification, such as doublet-triplet splitting  \cite{EW02}
due to discrete Wilson lines, are touched on in  subsection \ref{ssssgum}.

\subsection{The gauge group}
\label{group}

Physics beyond the standard model may involve extended gauge groups. In
particular, intersecting D6-brane models respectively D-brane models in
general involve $U(N)$ ($Sp(2N)$) groups for planes not parallel (parallel) to
orientifold planes. These may break to the standard model gauge group $G_{SM}
= SU(3)_C \times SU(2)_W \times U(1)_Y$ and additional $U(1)$ and non-Abelian
factors (the latter most commonly involves a quasi-hidden sector non-Abelian
group, as discussed in subsection \ref{susybreaking}). At intermediate stages
in all explicit $\mbb{Z}_2\times \mbb{Z}_2$ examples, $SU(3)_C$ is embedded
into a Pati-Salam $SU(4)$ in which lepton number is the fourth color
\cite{ps74}. In some cases, there is also an embedding of $SU(2)_W \times
U(1)_Y$ into $SU(2)_W \times SU(2)_R$. Here we focus on additional $U(1)$
factors, which frequently occur in intersecting D-brane constructions, as well
as other types of string constructions \cite{cl1996,plsusy04} and alternative
approaches to move beyond the Standard Model, such as dynamical symmetry
breaking \cite{hs2002} and Little Higgs models \cite{ahcg01,hlmw03}. Because
of their generality, extra $U(1)$ symmetries and their associated heavy $Z'$
bosons are probably the best motivated extension of the standard model after
supersymmetry. Experimental limits on an extra $Z'$ are very model dependent,
depending on the $Z'$ mass, gauge coupling, and couplings to the left and
right-handed quarks and leptons. However, typical limits from the combination
of $Z$-pole and other precision experiments and direct searches for $p \bar p
\rightarrow e^+e^-, \ \mu^+\mu^-$ are typically $M_{Z'} > 500-800$ GeV and the
$Z-Z'$ mixing less than a few $\times 10^{-3}$ (see, e.g., \cite{jk[pl04}).

There are several sources of extra $U(1)$ symmetries in intersecting D-brane
models. One is that stacks of branes not parallel to O6-planes yield $U(N)
\sim SU(N) \times U(1)$  groups, where the   $U(1)$'s are  typically
anomalous. Recall that the $U(1)$ anomalies are  canceled via a generalized
Green-Schwarz mechanism.
 As described in
subsections \ref{ssggsm} and \ref{ssnssm}, the  $Z'$ gauge bosons associated
with these  extra $U(1)$ factors will typically  acquire string-scale masses
by Chern-Simons terms, even if they are not anomalous \cite{imr01}. (Field
theoretic analogs have been studied recently in \cite{kn2004}.) However,
the
$U(1)$'s will survive as perturbative global symmetries of the theory, often  restricting
possible Yukawa couplings and/or leading to conserved baryon and lepton
numbers, stabilizing the proton and preventing Majorana neutrino masses
\cite{imr01}. For non-supersymmetric models with a TeV string scale this
implies new $Z'$ gauge bosons with masses generated without a Higgs mechanism.
Experimental constraints from their mixing with the $Z$ have been examined in
\cite{GIIQ02,DG02,DG02a}, where lower bounds on $M_{Z'}$ and therefore on the
string scale in the TeV range were obtained, somewhat more stringent than
typical bounds on $E_6$-motivated $Z'$ bosons.

Non-anomalous additional $U(1)$'s may arise from   the breaking of $SU(N)$
factors  by parallel splitting of  $U(N)$ branes, such as the extra
$U(1)_{B-L}$ emerging from $SU(4)$ in  \cite{csu01,csu01a} and other typical
Standard-like Model constructions;  or from the splitting of $Sp(2N)$ branes
parallel to O6-planes, such as the $Q_8-Q_{8'}$ in \cite{csu01,csu01a}. (See
\cite{clll04} for a general discussion.) As discussed in  \cite{cls02}, such
$U(1)$ factors need to be broken by the VEVs of standard model singlets
charged under the $U(1)$. The breaking could be at the TeV scale if it is
driven by the same type of terms which drive electroweak breaking
\cite{cdeel97}, or at a scale intermediate between the TeV and Planck scale
if it is along a D and (tree level) F-flat direction
\cite{cceel98}, provided there are appropriate Standard Model singlet fields.
In some cases, the only candidates are the bosonic partners of right-handed neutrinos
\cite{cls02}, and the needed couplings are not always present.

Experimental implications of a TeV-scale $Z'$ are significant.
These include the effects  at colliders of the $Z'$ and associated exotics
needed for anomaly cancellation  (see, e.g., \cite{jk[pl04}), and
the effects of the extended Higgs and neutralino sectors for
colliders and cosmology, commented on in subsection \ref{spectrum}.
The $Z'$ couplings are often family-nonuniversal in both intersecting
brane and heterotic constructions, implying flavor changing
neutral currents after fermion mixing is turned on (see, e.g.,  \cite{plmp00,ll2001, bcll04}).
These could be significant, e.g.,  for rare $B$, $K$, and $\mu$ decays.

\subsection{Gauge coupling unification}
\label{gaugeunif}

It is well known that the observed (properly normalized) low energy gauge couplings
$\alpha_1^{-1}\equiv \frac 3 5 \alpha_Y^{-1},$ $ \alpha_g^{-1},$ and $\alpha_s^{-1}$
associated respectively with $U(1)_Y$, $SU(2)_W$, and $SU(3)_C$
are
roughly consistent with gauge unification at a scale $M_U\sim 3 \times 10^{16}$ GeV
when the $\beta$ functions are calculated assuming the MSSM particle content (see, e.g.,
\cite{lp1995}). The value of $\alpha_s \sim 0.13$ predicted from
$\alpha$ and $\sin^2 \theta_W$ is slightly larger than the observed value ($\sim 0.12$),
even accounting for uncertainties in the sparticle spectrum, but could be due to
high scale threshold effects in traditional Grand Unified theories.
In heterotic string constructions one expects to maintain gauge unification
at the  string scale, which is typically an order of magnitude larger than
the apparent GUT scale $M_U$. However, the normalization of gauge couplings at $M_s$
is modified for higher Ka\v c-Moody embeddings, which are common for $U(1)_Y$
but not for $SU(3)_C \times SU(2)_W$. Also, most constructions involve additional
matter which can modify the $\beta$ functions. These two effects can each modify
the predicted $\alpha_s$ and $\log M_U$ by $O(1)$, compared to
the 10\% corrections that are needed, i.e., traditional gauge unification is lost
unless these two effects are absent or somehow compensate.

As discussed in subsection \ref{ssgc} traditional gauge unification is lost in
most intersecting D-brane constructions because the gauge coupling at the
string scale for each stack of D-branes depends on stack-dependent moduli,
i.e., on the volume of the three-cycle wrapped by the stack. (One exception
are supersymmetric Grand Unified Models \cite{csu01a,cps02}, in which the
standard model gauge factors all are derived from a single stack. A local 
construction \cite{ll03}, based on a three-family $U(3)^3$ sector also 
provides a gauge coupling unification of the three gauge factors at the 
string scale.)  Furthermore, as discussed in subsection \ref{spectrum} the constructions
(including the Grand Unified ones) typically involve exotic states that will
modify the running. One must therefore hope that these effects will somehow
compensate to yield the observed couplings.

One approach is to predict the low energy couplings (including those
for any additional gauge factors) for a given construction
in terms of the  spectrum.
This was done as a  function of $M_s$ and the volume moduli
in \cite{cls02} for the supersymmetric model \cite{csu01}
described in subsection \ref{sssz2}. It was found that the predicted couplings
were typically smaller than the observed ones due to the extra chiral matter.
The analysis was refined in  \cite{clw03}, in which it was shown
that due to gaugino condensation  the toroidal complex structure  moduli and
dilaton are fixed  (see also subsection \ref{susybreaking}).
One could then predict $\alpha_s^{-1}\sim 52.2$ and $\alpha(M_Z)^{-1}\sim 525$,
much larger than the observed values $\sim 8.5$ and $128$, due to the extra
chiral matter. The weak angle, which is a ratio, came out better, with the
predicted $\sin^2 \theta_W\sim 0.29$ not too far from the observed $0.23$.
While not successful, this illustrates the possibility that a more realistic
construction might lead to the observed couplings.

In a general D-brane construction there will not be a simple relation between
the three gauge couplings at $M_s$. However, it was observed in \cite{bls03}
that under certain circumstances there would be a tree-level relation
\begin{equation}
\alpha_1^{-1} = \frac 2 5 \alpha_s^{-1}+ \frac    3 5  \alpha_g^{-1},
\label{partialunif} \end{equation}
a special case of the canonical GUT relation in which all three are equal.
This could come about in models in which the weak
hypercharge satisfies the left-right symmetry relation $Q_Y =
\frac 1 2 (B-L) + Q_{3R}$, with the additional assumptions that
$U(1)_{B-L}$ derives from the same stack as $SU(3)_C$ (Pati-Salam embedding)
and that
there is a left-right symmetry that ensures the same coupling for
$Q_{3R}$ and $SU(2)_W$. It was shown in  \cite{bls03}
that if (\ref{partialunif}) holds, then from the observed low
energy couplings and from the contributions to the $\beta$ functions
from exotic matter one can predict the value of $M_s$ and the
volume moduli. In fact it turned out that an effective $\beta$ function
coefficient was always an even integer leading to a discrete set
of possible values for the string scale.
For example, for  no exotics one finds
$M_s \sim 2 \times 10^{16}$ GeV with volume radii
$R_s = 2.6/M_s$ and $R_W = 3.3/M_s$ for the $SU(3)_C$
and $SU(2)_W$ branes, respectively. The addition of exotic matter
can lead to very different $M_s$ and radii. Of course, there is no guarantee
that after stabilization $M_s$ and the moduli would actually take these values.
While the first assumption (on $U(1)_{B-L}$) is satisfied
by existing supersymmetric constructions because
 $SU(3)_C \times U(1)_{B-L} \subset SU(4)$, the second (on $Q_{3R}$ ) is
  not satisfied in most known constructions such as \cite{csu01,csu01a},
 which are not left-right symmetric. (Examples in which it does hold were
 given for the locally supersymmetric construction \cite{cim02d},
two
 models among supersymmetric
constructions in \cite{cll04} and a four-family model in \cite{clll04}.)
In \cite{dhs04a} the frequency of this relation was statistically investigated
in the ensemble of MSSM like Gepner model orientifolds. It was
found that for approximately 10\% of these models this relation
was satisfied (which could clearly be seen in the overall plot in \cite{dhs04a}).

\subsection{Yukawa couplings}
\label{Yukawaunif}

Yukawa couplings and the pattern of fermion masses and mixings are one of the
least understood aspects of nature. In the context of the Standard Model or
MSSM, or in simple grand unification extensions, it is often assumed that some
sort of additional family symmetry might lead to textures (hierarchies of
elements including zeroes)  in the fermion Yukawa matrices to explain the
observed patterns. The ratio $\tan \beta$ of the VEVs of the neutral Higgs
fields from the two Higgs doublets $H_u$ and $H_d$ of the MSSM may also play a
role. The recent observation of neutrino oscillations further complicates the
situation because of the possibility of Majorana masses.

In existing string constructions (including heterotic) the possible Yukawa and
other superpotential interactions are typically very much restricted by
additional symmetries (e.g., the perturbative
global symmetries that remain after anomalous
$U(1)$'s are broken by the Green-Schwarz mechanism), or by stringy selection
rules such as orbifold and orientifold projections. Such restrictions may be
weakened in more general constructions, but are an important feature of
existing examples, and they may lead, e.g, to texture zeros. For example, one
of the families of quark and lepton doublets in the supersymmetric  model
\cite{csu01} in subsection \ref{sssz2} has no Yukawa couplings due to the
$Q_2$ symmetry and remains massless, or the $SU(5)$ models described in
subsection \ref{ssssgum} have no  {\bf 10 10 5}$_H$ couplings due to the
$U(1)$ of $U(5)$. Many models (e.g., \cite{imr01,csu01}) have conserved $B$
and $L$ due to global and local $U(1)$'s, stabilizing the proton and
preventing Majorana neutrino masses. Similarly, in existing intersecting
D-brane models (unlike some heterotic constructions) $H_d$ and lepton doublets
are clearly distinguished even though they have the same Standard Model
quantum numbers because two of the relevant branes are distinct.

String constructions also allow natural mechanisms for hierarchies of Yukawa
couplings. For example, free fermionic models can lead to small effective
couplings from higher dimensional operators.
As described in subsection \ref{sssyc} intersecting D-brane constructions allow
for a geometrical origin of hierarchies, because allowed Yukawa couplings
are due to world-sheet instantons and are proportional to $\exp(-A)$,
where $A$ is the area of the triangle connecting the three intersecting branes.

Another aspect is that existing supersymmetric intersecting D-brane
constructions contain more than a single $H_{u,d}$ pair, as described in
subsection \ref{spectrum} (this is true for many heterotic constructions as
well), with each having different Yukawa matrices. Thus, hierarchies of their
VEVs could be an additional mechanism for achieving hierarchies of masses and
nontrivial mixings, and in generating otherwise vanishing masses. Of course,
the actual VEVs would depend on the details of how supersymmetry is broken. In
particular, in schemes of radiative electroweak breaking (in which negative
Higgs mass squares are generated from positive ones at a higher scale by
renormalization group running) there will be a strong tendency for only those
Higgs fields with large Yukawa couplings to actually acquire VEVs. There has
been relatively little phenomenological work on these sorts of extended Higgs
sectors.

In specific intersecting D-brane models on toroidal (orbifold)
compactifications, the Yukawa couplings often factorize in terms of the family
indices for the left and right handed fermions, e.g., the couplings
$h^k_{i,j}$ between $H_u^k$, $Q_i$ and $\bar U _j$ are proportional to
products $a^k_i b^k_j$. This can occur, for example, if the non-trivial
intersections for $Q_i$ and $\bar U _j$ occur in different two-tori
\cite{cim03} or if the orientifold and orbifold projections associate each
$\bar U _j$ with a distinct $H_u^j$ \cite{cls02a}. The factorization does {\em
not} hold in  more general examples (e.g., \cite{cll04}).  Factorization could
actually pose a problem for a construction with only a single pair of
$H_{u,d}$ doublets, because it allows only one massive state of each fermion
type ($u$-type, $d$-type, $e$-type). Some means must therefore be found to
populate other terms in the mass matrices. Possibilities include accepting
additional Higgs pairs, modifying the D-brane geometry, invoking (non-aligned)
four-point interactions in non-supersymmetric models with low $M_s$
\cite{als03}, or allowing for (non-aligned) supersymmetry breaking $A$ terms
(if allowed by the supersymmetry breaking mechanism)  \cite{als03}. There are
also potential problems with the minimal two-doublet structure if the
electroweak symmetry is promoted to $SU(2)_L \times SU(2)_R$, because the
$SU(2)_R$ symmetry would ensure equal Yukawa matrices for  the $u$ and $d$,
preventing a nontrivial CKM quark mixing matrix  \cite{cim02a,cim02d,cim03} .
(This is also one reason $SO(10)$ models require more than a single Higgs
multiplet coupling to fermions \cite{gut1980}.)

A more detailed analysis of the Yukawa couplings for the supersymmetric
multi-Higgs
model \cite{csu01} described
in subsection \ref{sssz2} was made in  \cite{cls02a}. It was shown
that for appropriate values of some (unknown) volume moduli one could obtain
nontrivial masses
and mixing for two families. Near the symmetric points (small splitting between stacks of branes)
one obtains the GUT-like result of similar $d$ and charged lepton mass matrices,
as well as similar $u$ and Dirac neutrino masses.
 The Dirac neutrino masses are problematic because
the model has no non-perturbative mechanism to generate Majorana masses for a
seesaw mechanism. The Yukawa structure for non-supersymmetric models with one
or two pairs of Higgs fields  were studied in \cite{imr01,als03}. A locally
supersymmetric model with a single Higgs pair (whose global embedding
was realized in \cite{dhs04,ms04,ms04a})
was considered in
\cite{cim02d,cim03}, where it was  emphasized that having only one massive
family is actually an excellent first approximation, since $m_t$, $m_b$, and
$m_\tau$ are much larger than the other generations.

\subsection{Flavor changing effects and proton decay}
\label{fcnc}

In the Standard Model there are no flavor changing neutral currents (FCNC)
 mediated by the $Z$, $\gamma$ or Higgs  at tree level,
and FCNC at loop level are  suppressed (the GIM mechanism). However, there are
enhanced FCNC effects in most extensions of the Standard Model,
including new loop effects
in supersymmetry and new interactions in dynamical symmetry breaking.
Similarly, the only sources of CP violation are the phases in the quark (and
lepton) mixings, possible neutrino Majorana phases, and a possible strong CP
parameter $\theta_{QCD}$. For small quark mixings, all but $\theta_{QCD}$ lead
to extremely small neutron, atomic,  and electric dipole moments (EDM), while
most extensions of the Standard Model lead to enhanced effects. Therefore,
experimental studies of rare decays and suppressed mixings, such as $\mu
\rightarrow 3 e$, $K_L-K_S$ mixing and rare $B$ decays, as well as refined
electric dipole moment experiments, are an excellent way to search for new
physics.

{\em FCNC.} There are a number of sources of FCNC in string constructions (in addition to
the standard sparticle loops in supersymmetric constructions). The tree level
calculation of the string four-point amplitudes \cite{ao03}  (see subsection
\ref{sstf}) produces  flavor changing four-point operators in the effective
action. For non-supersymmetric constructions with a low $M_s$, the analysis of
such operators  was carried out in \cite{ams03,als03}, where it was shown that
there could be significant effects from both Kaluza-Klein modes and stretched
heavy string modes. For example, Kaluza-Klein excitations couple
non-universally to states located at different positions, and therefore to
FCNC. The authors of \cite{ams03,als03} 
 studied the constraints on these operators from experimental founds
on FCNCs, EDMs, and supernova cooling by neutrino emission induced by
four-fermi operators, and showed that the FCNC especially severely restrict
the string scale to be higher than $\sim 10^4$ TeV.  This suggests that such
non-supersymmetric constructions have a severe fine-tuning problem, and make
it unlikely that other effects, such as the $U(1)$ gauge bosons which acquire
a string-scale mass by the Chern-Simons terms
 \cite{GIIQ02,DG02,DG02a} described above will be observable.

 A number of other (field theoretic) sources of FCNC may be expected from
 intersecting D-brane (and other) string constructions and may be observable in future
 experiments. The most promising are additional TeV-scale $U(1)$'s with family-nonuniversal
 couplings, as described in subsection \ref{group}; multiple Higgs doublets, for
 which the neutral components can mediate FCNC; or extended non-Abelian groups that
 can survive down to low energies, such as the embedding of $SU(2)_W$ into $Sp(6)$
 at $\sim 100$ TeV, leading to $K_L \rightarrow \mu^\pm e^\mp$ \cite{clll04}.

 {\em  CP violating phases.} The CP violating phases for supersymmetric constructions can appear in the Yukawa couplings which depend on the  VEVs of
  the (complex) K\"ahler moduli
   (for a detailed  discussion of this moduli dependence,  see \cite{cim03}).
   Another source of the  CP violating phases  can be complex   soft supersymmetry
  breaking  masses and the $\mu$ parameters; in intersecting D-brane constructions
 the  complex soft supersymmetry breaking mass parameters  are due to the complex VEVs of closed sector moduli,
 as discussed briefly in subsection \ref{susybreaking}.

{\em Strong CP problem.}
 In \cite{aiu02} a mechanism to solve the strong CP
problem was proposed, which could  have a realization within  
intersecting  D6-brane models.  This mechanism is reminiscent of the 
(chiral) anomaly inflow mechanism. Specifically, the proposal employs an 
additional bulk $U(1)_X$ gauge factor  under which quarks are not 
charged, and the flux associated with  the  NS-NS three-form field 
strength $H_3$.   The anomaly cancellation takes place
due to a Chern-Simons term of the Type IIA supergravity and  
terms in the expansion of the D6-brane  world-volume  Chern-Simons 
action.
A specific non-supersymmetric
intersecting D6-brane model, that explicitly realizes this mechanism 
was constructed in \cite{aiu02}. It remains an open problem to implement this mechanism
for  the supersymmetric intersecting
 D6-brane constructions with   supersymmetric $H_3$ fluxes.

{\em Proton decay.}
Supersymmetric Grand Unified theories \cite{pdg04} allow
proton decay by dimension 5 or dimension 6 operators (we assume that dimension
4 $R$ parity-violating terms that could lead to unacceptable rates are
absent). The dimension 5 operators  (via heavy colored fermion exchange) lead
to too rapid proton decay unless they are somehow forbidden, while the
dimension 6 operators from heavy gauge boson exchange typically lead to a
lifetime  of the order of $10^{36}$ yr, too long to observe in planned experiments
(the current limit of $\sim 4 \times 10^{33}$ yr for $p \rightarrow e^+
\pi^0$, which may be improved to $\sim 10^{35}$ yr).

The expectations for proton decay in supersymmetric  intersecting D-brane
constructions have been  studied  recently in  \cite{kw03}. In many
intersecting D-brane constructions baryon number is conserved perturbatively
and the proton is stable. However, the proton can decay in the Grand Unified
constructions described in \ref{ssssgum}.   The four-fermion contact operator
for {\bf 10}$^2${$ \overline{\rm \bf 10}$}$^2$ in intersecting D-brane $SU(5)$
models for the four states located at the same intersection (where there is no
suppression from area factors) was calculated in \cite{kw03} (see also
subsection \ref{sstf}). This operator has   an enhancement, relative to the
standard Grand Unified Models, due to the exchange of Kaluza-Klein excitations
of the color triplet gauge bosons, which leads to the decay amplitude $\propto
\alpha_{GUT}^{-1/3}$.  In order to further increase the decay amplitude,
 the  string coupling was taken to be ${\cal O}(1)$,
  thus leading to the M-theory on $G_2$ holonomy space (see
subsection \ref{ssg2}), and  the gauge coupling threshold corrections
 \cite{fw02}  were included.   However the final result did not have additional
large enhancement factors, suggesting a lifetime of around $10^{36}$ yr,
comparable to ordinary supersymmetric grand unification.

\subsection{Moduli Stabilization and Supersymmetry Breaking}
\label{susybreaking}
Intersecting D-brane constructions  on toroidal (orbifold) backgrounds possess
a large number of closed and open string sector moduli,
thus leading to a large  vacuum degeneracy.
In fact, the vacuum degeneracy problem is generic for supersymmetric string
constructions. As mentioned before, this problem has been addressed via
two mechanisms: (i)
implementation of the strong D-brane gauge dynamics that can lead to  gaugino
and matter condensations and generates a non-perturbative superpotential  for
the closed string sector moduli fields; (ii) introduction of supergravity
fluxes whose back-reaction introduces a moduli dependent potential.
 It is expected that in a realistic framework a combination of both
mechanisms will  play a role in
 obtaining string vacua with  (all) moduli stabilized, broken supersymmetry and
 potentially realistic cosmological constant. In the following we shall
 summarize the phenomenological implications, studied for these two
 mechanisms.

\subsubsection{Strong D-brane gauge dynamics}
\label{sssgd}

 Explicit supersymmetric intersecting D6-brane constructions typically
possess a quasi-hidden gauge sector  that   has a number of non-Abelian
confining gauge group factors, typically  with  $Sp(2N)$ gauge  symmetries.
The non-perturbative superpotential of the Veneziano-Yankielowicz-type
\cite{vy82,vty83} is  a sum of exponential factors (associated with each
confining gauge factor):
\begin{equation}
 W_{a}(U^i,S) =
\frac{\beta_a}{32 \pi^2} \frac{\Lambda^3}{e} \exp (\frac{8\pi^2}{\beta_a}
f_a(U^i,S)) \label{nw} \, , \end{equation}
 where the dynamically generated scale
  $\Lambda$ is roughly of the order of the string scale $M_s$, $\beta_a$ is the beta function of the
specific gauge group factor and  $f_a(U^i,S)$ denotes the corresponding gauge
kinetic function, which for intersecting D6-branes  depends on the complex
structure moduli $U^i$ and the dilaton field $S$. Eq.(\ref{nw})
accounts only for the leading instanton  contribution.  One should also point
out that for a specific number of ``flavor'' (matter) $N_f$ and ``color''
(gauge) $N_c$ degrees of freedom  there are 
subtleties; e.g.,  for  $Sp(2N_c$) gauge factors, can lead
to the quantum lift of the moduli space ($N_f=N_c+1$) or absence of the
non-perturbatively generated global superpotential ($N_f>N_c+2$). (For a
review see, e.g., \cite{is95} and references therein; for the implementation
of strong gauge dynamics in  the effective actions from heterotic strings, see
\cite{kl93}.) Classes of semi-realistic supersymmetric intersecting D6-branes
constructions, e.g., \cite{cp03a,cll04}, have the property that the hidden
sector gauge group factors satisfy $N_f<N_c+1$, resulting in  confining 
infrared
dynamics and the non-perturbative superpotential of the type (\ref{nw}).

For toroidal (orbifold) compactifications, as discussed in subsection
 \ref{ssgc},
the tree-level gauge kinetic function $f_a(U^i,S)$ (\ref{tgc})  depends on the
 dilaton $S$ and three toroidal complex structure moduli $U^i$
 (some of  the  toroidal complex structure moduli
 are fixed by the supersymmetry
 constraints in the D6-brane sector),
 and the specific  wrapping numbers  $(n^i_a,{\widetilde m}^i_a)$ of the
 three-cycle $\pi_a$, wrapped by a  stack of $N_a$ D6-branes.
 For a specific supersymmetric
 semi-realistic construction \cite{csu01,csu01a} the  non-perturbative
 superpotential (\ref{nw}),  associated with the
 confining $Sp(2)\times Sp(2)\times Sp(4)$ sector, resulted \cite{clw03} in  the minimum
of the potential
 that stabilized the
  remaining toroidal complex structure modulus $U$ and the dilaton $S$, and broke supersymmetry.
It would also be interesting to implement the threshold corrections to the
gauge kinetic function \cite{ls03} as discussed in subsection \ref{ssgc}.
For ${\cal N}=2$ sectors these
corrections  depend  also on  toroidal K\"ahler moduli, and thus  the
non-perturbative superpotential (\ref{nw})  could in principle allow for the
stabilization of the toroidal K\"ahler  moduli as well.

 When supersymmetry is broken due to such a  non-perturbative superpotential
the gaugino masses $m_{\lambda_a}$
 can  be determined in terms of F-breaking terms associated with $S$ and $U^i$ moduli
 directions:
 \begin{equation}
m_{\lambda_a}=(\partial_{\phi^i}\,f_a(\Phi^i))K^{\Phi^i \, {\bar\Phi^j}} {\bar
F_{\bar \Phi^j}}\, . \end{equation}
 Here $K^{\Phi^i \, {\bar\Phi^j}}$ is the
inverse of the K\"ahler metric of the moduli $\Phi^i$,
and $F_{\Phi^j}$ are the F-breaking-terms for the
moduli $\Phi^j=\{S,\ U^i\}$.
 Unlike the heterotic constructions and simple Grand Unified theories,
the gaugino masses and gauge couplings at the string scale depend on more than
one modulus, i.e.,  $S$ and $U_i$, which in general have complex VEVs,   and
thus lead to non-universal and  complex (indicating significant $CP$-violating
phases) gaugino masses. Unfortunately for the specific model, studied in
\cite{clw03},  these masses were too heavy, i.e., ${\cal O}(10^8)$ GeV.

The study of soft supersymmetry breaking parameters of the charged matter
sector requires detailed information on the moduli dependence of the leading
term in the K\"ahler potential  for the charged matter; this K\"ahler
potential  was recently determined  in \cite{lmrs04,cp03} and discussed in
subsection \ref{ssskp} (specifically, see  eq. (\ref{Kahler})). Unfortunately,
for the specific example studied in \cite{clw03} the minimum of the
non-perturbative superpotential produced a large negative cosmological
constant, and thus these vacua do not provide realistic backgrounds for a
detailed study of the soft supersymmetry breaking parameters of the charged
matter sector. However, one  can assume that the non-perturbative mechanism
for supersymmetry breaking does not introduce a large cosmological constant,
and then  one can parameterize such soft masses via $F_{\Phi_i}$-breaking
terms associated with  moduli  $\Phi_i$, by employing the standard
supergravity  techniques.  Such a study was recently performed in
\cite{kklw04}. (For an earlier work see \cite{kn03}.) In the regime where the
$F_{U_i}$-breaking terms, associated with the $U_i$ moduli, are dominant, the
mass parameters do not depend on the Yukawa couplings, and have a pattern
different from the heterotic string.

In principle the  strong gauge dynamics can also lead to composite
(baryon-type) states whose constituents  include  states that are chiral
exotics, i.e., states charged both under  the Standard Model gauge  factors and
the hidden strong gauge sectors. This scenario could provide another mechanism
to remove chiral exotics from the light  spectrum (see \cite{cls02}).

\subsubsection{Supergravity fluxes}

Supergravity fluxes  provide another mechanism  to stabilize the
compactification moduli. The supersymmetric flux compactifications are better
understood on the Type IIB side (for details see section \ref{ssbf}).
Semi-realistic constructions of Type IIB vacua consist of the magnetized
D-brane sector (T-dual to the intersecting D6-branes) and the $G_3$ fluxes
stabilizing  the
toroidal complex structure moduli (in the T-dual picture K\"ahler moduli)  and the
dilaton-axion field.

Typical semi-realistic examples have fluxes that break supersymmetry via
a $(0,3)$ component of $G_3$, and
recent phenomenological studies focused on the implied generation  of
 soft  supersymmetry breaking terms in the low energy effective action
\cite{ciu03,ggjl03,lrs04,lrs04a,ciu04,IBANEZ04,msw04,fi04,kklw04}.
 These terms  have been
derived by employing two  complementary approaches:
\begin{itemize}
\item{ The soft supersymmetry
breaking mass terms due to  fluxes were obtained  by expanding the resulting
Dirac-Born-Infeld action for the D3 and D7 branes
\cite{ciu03,ggjl03,ciu04} to the  lowest order in the coordinates
transverse to the D-brane world-volume.}
\item{ Employing the standard supergravity formalism, one can parameterize
the soft supersymmetry breaking  terms  via   the supersymmetry breaking VEVs
 of the auxiliary F and D components of the
chiral and vector supermultiplets \cite{lrs04,lrs04a}.}
\end{itemize}
Both approaches are expected to be equivalent. A third approach using
the F-theory description of a certain orientifold has been pursued
in \cite{lmrs05}.

We have given in section \ref{ssbf} a heuristic argument why
such soft terms
are generated.
In the following we summarize the specific results.
For the imaginary
self-dual $G_3$, the  soft supersymmetry breaking mass terms are absent for
the matter associated with the open string states on the D3-branes. However,
for the anti-D3-branes these masses  are non-vanishing, and specifically they
stabilize the open-string modulus associated with the position of the
anti-D3-brane \cite{ciu03,lrs04}. This point also plays a very important
role in getting de-Sitter vacua via the KKLT construction \cite{kklt03}.

On the other hand, such mass terms  for the
open string states on D7-branes  are due to the  non-supersymmetric
$(0,3)$-components of $G_3$ fluxes \cite{ciu04,lrs04a}. Interestingly, the
supersymmetric  $(2,1)$-components of $G_3$ fluxes can induce
superpotential mass terms for the D7-brane moduli, including those
associated with D7-brane ``intersections'' \cite{ciu04}, thus providing a
stabilization mechanism for them.
Assuming a homogeneous flux,
the scale of such mass terms
is of
 the order
${M_s^2}\over {M_{pl}}$. For the supersymmetry breaking masses to be of
the TeV scale, this
implies  that the string  scale is   in the intermediate
regime.
This is reminiscent of the gravity mediated supersymmetry breaking
mechanism.
More generally,
such mass terms measure
the {\it local} flux density and so $M_s$ that appears in the above
estimate should be the {\it local} string scale
which can 
in principle be much smaller because
of the non-trivial warp factor.

\subsection{Cosmological aspects}

The main focus of this review has been the particle physics
aspects of intersecting D-brane models.
For completeness, however, let us briefly mention
some cosmological aspects of this scenario as well.
A comprehensive overview of
 string/brane cosmology
is beyond the scope of this review. Here, we shall only sketch some highlights
of this subject that are particularly relevant to intersecting D-brane worlds.
For details and references, we refer the readers to some excellent reviews
\cite{Quevedo,Danielsson,Polchinski}.

There has been widespread hope that string theory may provide a microscopic
origin for inflation. The discovery of D-branes has opened up several new
possibilities. In this review, we have focused on D-brane models that preserve
${\cal N}=1$ supersymmetry, for otherwise the D-brane configurations are
generically unstable with too short a life-time that would spell disaster for
particle physics today. However, in the early universe, the initial
configuration of D-branes is not necessarily perfectly stable. Instead the
D-branes could intersect at non-supersymmetric angles, or there could be
additional pairs of branes and anti-branes separated in the compact
dimensions. These instabilities drive the system of D-branes to a neighboring
stable configuration, so we can think of the supersymmetric models that we
have discussed at length in this review as the endpoints of such dynamical
processes. In fact, a natural candidate for the inflaton field in this
scenario is the open string mode whose VEV describes the inter-brane
separation \cite{dt99}. The dynamics of inflation is therefore governed by the
interaction between D-branes. This idea of brane inflation \cite{dt99} has
been applied to construct inflationary models arising from the collision of
branes and anti-branes \cite{dt99,Burgess:2001fx,dss01,st01,kklmmt03}, as well
as branes intersecting at angles \cite{grz01,bklo02,gz02}. In particular,
\cite{kklmmt03}, which is by far the most detailed model of inflation from
string theory, demonstrated that the brane inflation proposal can be
implemented in a string model where the geometric moduli are stabilized by the
background fluxes (a concrete mechanism that we discussed in section
\ref{ssbf}).

Interestingly, the cosmic string network produced at the end of D-brane
inflation offers an exciting opportunity to test stringy physics from
cosmological observations (see, e.g.,\cite{Polchinski,Shiu} for some reviews
and references).
Towards the end of brane inflation, the inflaton potential becomes tachyonic.
The condensation of this complex tachyon mode results in the formation of
cosmic strings, rather than other cosmological defects such as monopoles or
domain walls.
Finally, in addition to cosmic D-strings, there could in general 
be stable D-branes carrying K-theory charges in the 
intersecting D-brane models discussed here. They could be
interesting candidates for superheavy
dark matter \cite{sw03}.

\section{CONCLUSIONS AND OUTLOOK}

In this article we have provided a pedagogical introduction
to string theoretic  intersecting D-brane models, which we
hope suits the need of students to have a comprehensive though
not too technical guideline for this topic.
In the second part we have tried to briefly review much of the work
on intersecting D-brane models carried out so far, including
an overview on model building attempts including
recent flux compactifications as well as
on the structure of the low-energy effective action.
The latter of course is very important for concrete
phenomenological applications of these models.

During the short history of intersecting D-brane constructions,
it has happened  several times that
new momentum was brought into the field from other
branches of string theory research like M-theory
compactifications on $G_2$ manifolds or flux compactifications.
Clearly, all these model building schemes are intimately
related. After more than four years of intense research, it
has become clear that intersecting D-brane models
provide  a general phenomenologically
appealing class of string compactifications, which can
also  be considered as honest  string theory realizations
of some of the ideas concerned with  more phenomenologically motivated
brane world models.

Even though we have a nice geometric framework, we are still lacking a
completely convincing model realizing the MSSM. One can find isolated
mechanisms for realizing  most of the features of the Standard Model like
family replication, hierarchical Yukawa couplings, absence of extra gauge
symmetries and vector-like matter etc., but all concrete models studied so
far do  not realize all Standard Model properties at the same time. They
either have extra chiral exotic matter as   is typical for supersymmetric
constructions,  or the models fail at the level of couplings, such as gauge
and Yukawa couplings.
Of course only a  very few  classes of models, primarily  based on toroidal
orbifolds,   were constructed and even fewer were studied  in detail. In
addition,  the techniques are not yet available to study more general
intersecting D-brane models on say generic smooth Calabi-Yau spaces. In fact
it seems to be the case that, for instance, the notorious appearance of extra
vector-like matter is related to the fact that we are only considering models
at very special, highly symmetric points in moduli space like orbifold or
Gepner points. In view of the impression that the finer details of the
Standard Model are far from being very natural, there is no guarantee that
nature has finally stabilized in a string vacuum which is highly symmetric and
treatable with the simple methods developed so far. Therefore, it would be
interesting to develop the tools to study more generic intersecting D-brane
models.

Alternatively, it is entirely possible that physics at the TeV scale
is richer than the MSSM, and that some of the features found
in existing constructions, such as extended gauge symmetries, extra
chiral matter, and flavor changing neutral
currents really exist. The LHC and future experimental probes
are eagerly awaited to refine the target of our theoretical investigations.

As reviewed in this article, concerning the low energy effective field theory
considerable  progress has been made in computing for example Yukawa couplings,
the  K\"ahler potential
or the resulting soft supersymmetry breaking terms
for very simple toroidal backgrounds. However, much more work is needed
to derive similar results for more general backgrounds.

As should be clear, for each Calabi-Yau manifold there does exist
a plethora of consistent intersecting D-brane models.
In view of these, one might ask whether there does exist
any chance  to find the/a realistic string vacuum.
This picture becomes even more severe when one also takes
the so-called landscape of flux compactifications into account.
It was proposed that complementary to a model by model search,
one 
could study the statistical distribution of string theory
vacua \cite{MD03} (see \cite{bghlw04} for a statistical  analysis
of intersecting D-branes)
 to obtain an estimate of the chances of finding
a realistic model  and maybe in which region of the
parameter space one should look.

For the moment we can only hope that continuous work on both
approaches -  the model by model search and the statistical analysis - will
eventually lead us to a realistic string model, from which,
once the background is fixed, all features of the low-energy
effective theory can be derived. However, whether such a model  is
in any sense unique is not guaranteed, as we will always
measure the physical parameters with some finite accuracy.
Having one string model which describes our world
within the accuracy of our measurements would  nevertheless
be considered a milestone in our understanding of nature.

\vskip 2cm
\noindent

{\bf Acknowledgements:} R.B. would like to thank PPARC for financial support during the
first half of this project. The research was supported in part by
Department of Energy Grant
DOE-EY-76-02-3071 (M.C. and P.L.), Fay R. and Eugene L. Langberg 
Endowed Chair  (M.C.), National Science Foundation  Grant
INT02-03585 (M.C. and G.S.), National Science Foundation  CAREER Award 
PHY-0348093 (G.S.), Department of Energy  Grant DE-FG-02-95ER40896 (G.S.) 
and  a Research  Innovation Award from 
Research Corporation (G.S.).
Part of this review has been written
at DAMTP (University of Cambridge) and R.B., M.C. and G.S. would like to thank
DAMTP for hospitality. 
R.B. also thanks the University of Pennsylvania for hospitality.
G.S. thanks the Perimeter Institute for Theoretical
Physics for hospitality
during the final stage of writing this review.
We would like to thank G. Honecker, F. Marchesano, T. Liu, and T. Weigand 
for useful comments about the manuscript,
and our various collaborators, including C. Angelantonj, V. Braun, 
J.P. Conlon, F. Gmeiner, L. G\"orlich, B. Greene, G. Honecker,
B. K\"ors, T. Li, T. Liu, D. L\"ust, F. Marchesano,
T. Ott, I. Papadimitriou, K. Schalm,
S. Stieberger, K. Suruliz, T. Taylor, H. Tye, A. Uranga, L.-T. Wang,
and T. Weigand.

\clearpage \nocite{*}
\bibliography{rev}
\bibliographystyle{utphys}

\end{document}